\documentclass[twocolumn,preprintnumbers,prd,superscriptaddress,nofootinbib,floatfix]{revtex4-2}
\usepackage{graphicx}
\usepackage{amsmath,amsfonts,amssymb}
\usepackage{physics}
\usepackage{tabularray}
\usepackage{threeparttable}
\usepackage{comment}
\usepackage{float}
\usepackage[dvipsnames]{xcolor}
\usepackage{bm}
\usepackage{soul} 
\usepackage{natbib}

\usepackage[colorlinks=true,linkcolor=blue,urlcolor=blue,citecolor=blue,anchorcolor=blue]{hyperref}
\usepackage[capitalize]{cleveref}

\UseTblrLibrary{booktabs}
\begin{document}
\title{Revisiting CPL with sign-switching density: to cross or not to cross the NECB}

\author{Mine G\"{o}k\c{c}en}
\email{gokcen23@itu.edu.tr}
\affiliation{Department of Physics, Istanbul Technical University, Maslak 34469 Istanbul, T\"{u}rkiye}

\author{\"{O}zg\"{u}r Akarsu}
\email{akarsuo@itu.edu.tr}
\affiliation{Department of Physics, Istanbul Technical University, Maslak 34469 Istanbul, T\"{u}rkiye}

\author{Eleonora Di Valentino}
\email{e.divalentino@sheffield.ac.uk}
\affiliation{School of Mathematical and Physical Sciences, University of Sheffield, Hounsfield Road, Sheffield S3 7RH, United Kingdom}

\begin{abstract}
Recent DESI DR2 BAO measurements, when combined with CMB and SNeIa data, exhibit a $3.2\sigma$--$3.4\sigma$
preference for dynamical dark energy described by the Chevallier--Polarski--Linder (CPL)-parametrized
equation of state.
A particularly striking feature of these reconstructions is an apparent transition from an early-time
phantom-like regime to a late-time quintessence-like behavior, whose theoretical realization is highly
nontrivial.
For positive-definite dark energy densities, this transition is often phrased as a crossing of the
phantom divide line (PDL) at $w(a)=-1$.
Allowing the dark energy density to become negative, however, renders the PDL (in the sense of
$w(a)=-1$) non-diagnostic as a global separator: the physically meaningful criterion is instead the null
energy condition boundary (NECB), $\rho_{\rm DE}+p_{\rm DE}=0$.
We therefore test whether the data-driven preference for NECB-crossing in CPL reconstructions persists
once alternative realizations of phantom behavior are admitted, specifically through sign-switching dark
energy densities.
To this end, we introduce and constrain two controlled phenomenological extensions of the CPL framework
featuring a negative dark energy phase in the past.
In the CPL$\to-\Lambda$ model, the switching epoch is tied to the CPL-inferred NECB-crossing scale factor,
yielding an early-time negative cosmological-constant phase, while the post-switch evolution follows the
CPL branch.
In the sCPL model, the CPL equation of state is maintained at all times, while the sign switch in the
energy density occurs at an independent transition redshift.
We find that late-time BAO and SNeIa data drive the negative-density phase beyond their effective redshift
coverage, and that this requirement is the primary driver of the inferred parameter behavior.
While both models are statistically disfavored relative to the baseline CPL, admitting a negative dark
energy phase generally reduces the significance of deviations from a cosmological constant.
\end{abstract}

\maketitle


\section{Introduction}
\label{sec:intro}

The concordance model of cosmology, the $\Lambda$CDM model~\cite{Peebles:2002gy, Copeland:2006wr}, has been subjected to increasingly rigorous scrutiny following the emergence of cosmological tensions~\cite{Verde:2019ivm,DiValentino:2020zio,DiValentino:2021izs,Perivolaropoulos:2021jda,Schoneberg:2021qvd,Shah:2021onj,Abdalla:2022yfr,DiValentino:2022fjm,Kamionkowski:2022pkx,Giare:2023xoc,Hu:2023jqc,Verde:2023lmm,Akarsu:2024qiq,DiValentino:2024yew,CosmoVerseNetwork:2025alb,Ong:2025cwv} revealed by the new generation of high-precision observations~\cite{Planck:2018nkj, Planck:2018vyg, AtacamaCosmologyTelescope:2025blo,SPT-3G:2025bzu, eBOSS:2020yzd, DESI:2025zgx,Scolnic:2021amr,Brout:2022vxf,Rubin:2023jdq, Popovic:2025glk, DES:2025sig}. Over the past decade, a wealth of increasingly precise measurements from a wide range of cosmological probes has become available, opening new avenues for cosmological inference and enabling nearly model-independent determinations of key cosmological parameters. In particular, the growing discrepancy between the value of the Hubble constant $H_0$ inferred from cosmic microwave background (CMB) observations~\cite{Planck:2018vyg,SPT-3G:2025bzu} and that obtained from direct local distance measurements~\cite{Freedman:2020dne,Birrer:2020tax,Riess:2021jrx,Anderson:2023aga,Scolnic:2023mrv,Jones:2022mvo,Anand:2021sum,Freedman:2021ahq,Uddin:2023iob,Huang:2023frr,Li:2024yoe,Pesce:2020xfe,Kourkchi:2020iyz,Schombert:2020pxm,Blakeslee:2021rqi,deJaeger:2022lit,Murakami:2023xuy,Breuval:2024lsv,Freedman:2024eph,Riess:2024vfa,Vogl:2024bum,Scolnic:2024hbh,Said:2024pwm,Boubel:2024cqw,Scolnic:2024oth,Li:2025ife,Jensen:2025aai,Riess:2025chq,Benisty:2025tct,Newman:2025gwg,Stiskalek:2025ktq,H0DN:2025lyy,Agrawal:2025tuv,Bhardwaj:2025kbw} has reached a statistically significant level above $7\sigma$~\cite{H0DN:2025lyy}. As a result, the validity of the $\Lambda$CDM framework has begun to be questioned to an unprecedented extent. 

A wide variety of potential solutions to the $H_0$ tension have been proposed over the past decade. These include detailed investigations of possible observational systematics~\cite{Efstathiou:2020wxn,Mortsell:2021nzg,Mortsell:2021tcx,Riess:2021jrx,Sharon:2023ioz,Murakami:2023xuy,Riess:2023bfx,Bhardwaj:2023mau,Brout:2023wol,Dwomoh:2023bro,Uddin:2023iob,Riess:2024ohe,Freedman:2024eph,Riess:2024vfa}, modifications of gravity~\cite{DiValentino:2015bja,Zumalacarregui:2020cjh,Odintsov:2020qzd,Adi:2020qqf,DeFelice:2020cpt,Pogosian:2021mcs,CANTATA:2021asi,Schiavone:2022wvq,Ishak:2024jhs,Specogna:2023nkq,Specogna:2024euz,AtacamaCosmologyTelescope:2025nti,Giare:2025ath,Tiwari:2023jle,Hogas:2023pjz,Wen:2023wes,Pitrou:2023swx,Montani:2024pou,Dwivedi:2024okk,Akarsu:2024qsi,Akarsu:2024nas,Hogas:2025ahb}, extensions of the standard cosmological parameter space~\cite{DiValentino:2016hlg}, scenarios involving vacuum phase transitions~\cite{DiValentino:2017rcr}, and a broad class of models based on evolving or dynamical dark energy~\cite{Dutta:2018vmq,vonMarttens:2019ixw,Akarsu:2019hmw,DiValentino:2020naf,DiValentino:2020vnx,Yang:2021flj,DiValentino:2021rjj,Heisenberg:2022lob,Akarsu:2022lhx,Giare:2023xoc,Adil:2023exv,Gomez-Valent:2023uof,Lapi:2023plb,Krolewski:2024jwj,Bousis:2024rnb,Tang:2024gtq,Jiang:2024xnu,Manoharan:2024thb,Specogna:2025guo,Ozulker:2025ehg,Lee:2025pzo}. 
Additional proposals have focused on early-time modifications of the expansion history, often referred to as early dark energy or early-universe solutions~\cite{Poulin:2018cxd,Smith:2019ihp,Niedermann:2019olb,Krishnan:2020obg,Schoneberg:2021qvd,Ye:2021iwa,Poulin:2021bjr,Niedermann:2021vgd,deSouza:2023sqp,Poulin:2023lkg,Cruz:2023lmn,Niedermann:2023ssr,Vagnozzi:2023nrq,Efstathiou:2023fbn,Simon:2024jmu,Cervantes-Cota:2023wet,Garny:2024ums,Giare:2024akf,Giare:2024syw,Poulin:2024ken,Pedrotti:2024kpn,Kochappan:2024jyf,Poulin:2025nfb,Smith:2025zsg}, as well as late-time or local transitions in the expansion rate~\cite{DiValentino:2019exe,Alestas:2021luu,Ruchika:2023ugh,Frion:2023xwq,Ruchika:2024ymt}. Other approaches have explored models with negative or sign-switching dark energy densities~\cite{Akarsu:2019hmw,Visinelli:2019qqu,Ye:2020btb,Calderon:2020hoc,Akarsu:2021fol,Sen:2021wld,DiGennaro:2022ykp,Akarsu:2022typ,Ong:2022wrs,Akarsu:2023mfb,Anchordoqui:2023woo,Adil:2023exv,Akarsu:2024qsi,Halder:2024uao,Anchordoqui:2024gfa,Akarsu:2024eoo,Yadav:2024duq,Paraskevas:2024ytz,Gomez-Valent:2024tdb,Toda:2024ncp,Gomez-Valent:2024ejh,Akarsu:2025gwi,Souza:2024qwd,Soriano:2025gxd,Akarsu:2025ijk,Escamilla:2025imi,Bouhmadi-Lopez:2025ggl,Ghafari:2025eql, Ibarra-Uriondo:2026zbp}, emerging dark energy scenarios~\cite{Pan:2019hac,Yang:2020zuk,Yang:2021eud}, dark matter or neutrino-based solutions~\cite{DiValentino:2017oaw,Anchordoqui:2022gmw,Pan:2023frx,Allali:2024anb,Co:2024oek,Aboubrahim:2024spa}, and possible revisions of recombination-era physics~\cite{Hart:2017ndk,Hart:2019dxi,Sekiguchi:2020teg,Hart:2021kad,Lee:2022gzh,Chluba:2023xqj,Greene:2023cro,Greene:2024qis,Baryakhtar:2024rky,Seto:2024cgo,Mirpoorian:2024fka,Lynch:2024hzh,Toda:2024ncp,Schoneberg:2024ynd,Smith:2025uaq,GarciaEscudero:2025lef,Toda:2025kcq}. 

Among these possibilities, late-time departures from the $\Lambda$CDM model have attracted particular attention. This is largely due to the many remaining uncertainties surrounding the dark sector and the dominant role played by dark energy (DE) in shaping the recent expansion history of the Universe. Independent of the current tensions, both parametric and non-parametric reconstructions of the DE evolution~\cite{Cooray:1999da,Efstathiou:1999tm,Wetterich:2004pv,Corasaniti:2002vg,Bassett:2002fe,Alam:2004jy,Linder:2005ne,Melchiorri:2006jy,Zhang:2006em,Li:2012vn,Novosyadlyj:2013nya,Rezaei:2017yyj,Wang:2017lai,Rezaei:2019hvb,Yang:2021flj,Sharma:2022ifr,vonMarttens:2022xyr,Yao:2022jrw,Feng:2004ff,Hannestad:2004cb,Xia:2004rw,Gong:2005de,Jassal:2005qc,Nesseris:2005ur,Liu:2008vy,Barboza:2008rh,Barboza:2009ks,Ma:2011nc,Sendra:2011pt,Feng:2011zzo,Barboza:2011gd,DeFelice:2012vd,Feng:2012gf,Wei:2013jya,Magana:2014voa,Akarsu:2015yea,Pan:2016jli,DiValentino:2016hlg,Nunes:2016plz,Nunes:2016drj,Magana:2017usz,Yang:2017alx,Pan:2017zoh,Panotopoulos:2018sso,Yang:2018qmz,Jaime:2018ftn,Das:2017gjj,Yang:2018prh,Li:2019yem,Yang:2019jwn,Pan:2019hac,Tamayo:2019gqj,Pan:2019brc,DiValentino:2020naf,Rezaei:2020mrj,Perkovic:2020mph,Banihashemi:2020wtb,Jaber-Bravo:2019nrk,Benaoum:2020qsi,Yang:2021eud,Jaber:2021hho,Alestas:2021luu,Yang:2022klj,Escudero:2022rbq,Castillo-Santos:2022yoi,Yang:2022kho,Akarsu:2022lhx,Dahmani:2023bsb,Escamilla:2023oce,Rezaei:2023xkj,Adil:2023exv,Escamilla:2023shf,LozanoTorres:2024tnt,Singh:2023ryd,Giare:2024akf,Rezaei:2024vtg,Reyhani:2024cnr,Escamilla:2024ahl,Akarsu:2026anp} have long served as important tools for probing deviations from a cosmological constant. 
In this context, the Chevallier-Polarski-Linder (CPL) parametrization~\cite{Chevallier:2000qy,Linder:2002et} of the dark energy equation of state has become one of the most widely adopted frameworks. Its popularity stems from its simplicity, flexibility, and ability to capture a broad range of time-varying dark energy behaviors within a minimal two-parameter description.

The latest Data Release 2 (DR2) of the Dark Energy Spectroscopic Instrument (DESI) collaboration, based on baryon acoustic oscillation (BAO) measurements~\cite{DESI:2025zgx}, has significantly reshaped the landscape of dynamical dark energy (DDE) studies. When combined with CMB anisotropy data and type Ia supernova (SNeIa) measurements, the DESI BAO results exhibit a greater-than $3\sigma$ preference for a cold dark matter cosmology with CPL-parametrized dark energy over the standard $\Lambda$CDM model~\cite{Hoyt:2026fve,DES:2025sig}. This preference has been confirmed across a variety of data combinations and parametrizations~\cite{DESI:2025fii,DESI:2024mwx,DESI:2025zgx,Cortes:2024lgw,Shlivko:2024llw,Luongo:2024fww,Gialamas:2024lyw,Dinda:2024kjf,Najafi:2024qzm,Wang:2024dka,Ye:2024ywg,Tada:2024znt,Carloni:2024zpl,Chan-GyungPark:2024mlx,DESI:2024kob,Bhattacharya:2024hep,Ramadan:2024kmn,Pourojaghi:2024tmw,Giare:2024gpk,Reboucas:2024smm,Giare:2024ocw,Chan-GyungPark:2024brx,Li:2024qus,Jiang:2024xnu,RoyChoudhury:2024wri,Escamilla:2024ahl,Sabogal:2024qxs,Li:2025cxn,Wolf:2025jlc,Shajib:2025tpd,Giare:2025pzu,Chaussidon:2025npr,Kessler:2025kju,Pang:2025lvh,Roy:2024kni,RoyChoudhury:2025dhe,Paliathanasis:2025cuc,Scherer:2025esj,Giare:2024oil,Liu:2025mub,Teixeira:2025czm,Santos:2025wiv,Specogna:2025guo,Sabogal:2025jbo,Cheng:2025lod,Herold:2025hkb,Cheng:2025hug,Ozulker:2025ehg,Gialamas:2025pwv,Lee:2025pzo,Ormondroyd:2025iaf,Silva:2025twg,Ishak:2025cay,Fazzari:2025lzd,RoyChoudhury:2025iis,Smith:2025icl,Li:2025vuh,Zhang:2025lam,Cheng:2025yue,Xu:2026sbw}. 
At the same time, these results tend to exacerbate the existing $H_0$ tension, further motivating the exploration of alternative theoretical frameworks. In particular, constructing dark energy models capable of reproducing the specific phenomenological features favored by DESI BAO presents a new set of theoretical challenges. One of the most notable characteristics emerging from CPL-based reconstructions is a transition
from an early-time phantom-like regime to a late-time quintessence-like behavior, an evolution
that is highly nontrivial to realize within consistent dark energy theories.
In the standard setting of positive-definite dark energy density, this is
often phrased as a crossing of the phantom divide line (PDL) at $w_{\rm DE}=-1$, or simply
``phantom crossing''.
In that restricted case ($\rho_{\rm DE}>0$, as is the case for the standard CPL density by
construction), the null energy condition boundary (NECB) $\rho_{\rm DE}+p_{\rm DE}=0$ is equivalent to $w_{\rm DE}=-1$.
Once $\rho_{\rm DE}$ is allowed to change sign, however, the PDL in the sense of $w_{\rm DE}=-1$
ceases to be a global separator: the physical regime is determined by the sign of
$\rho_{\rm DE}+p_{\rm DE}$ (or equivalently $(1+w_{\rm DE})\rho_{\rm DE}$), and $w_{\rm DE}$ itself
can become non-diagnostic if and when $\rho_{\rm DE}$ crosses zero.
Accordingly, throughout this work we characterize regime changes in terms of the NECB rather
than $w_{\rm DE}$ alone~\cite{Carroll_2003,Adil:2023exv,Caldwell:2025inn,Akarsu:2026anp}.

The NECB-crossing behavior inferred within the CPL parametrization was recently examined in Ref.~\cite{Ozulker:2025ehg}, where two restricted versions of CPL dark energy were introduced by enforcing a $\Lambda$-like equation of state either before or after reaching the NECB. These constructions were designed to test whether the apparent crossing preferred by the data reflects a genuine physical feature or instead arises as a reconstruction artifact of the linear CPL form. While strong empirical support for an effective NECB-crossing was reported, that analysis was limited to scenarios in which the dark energy density remains positive definite.
Motivated by this limitation, the purpose of the present work is to revisit the NECB-crossing preference in a more general setting, allowing the dark energy density itself to change sign. We therefore introduce two phenomenological extensions of CPL-like dark energy that incorporate sign-switching energy densities, remaining positive today while becoming negative in the past. This framework enables a direct test of whether the data-driven preference for NECB-crossing persists once alternative realizations of phantom behavior are admitted, independently of the linear CPL parametrization. Our constructions are further informed by sign-switching scenarios such as $\Lambda_{\rm s}$CDM~\cite{Akarsu:2019hmw,Akarsu:2021fol,Akarsu:2022typ,Akarsu:2023mfb}, which motivate exploring a negative dark energy density phase that can become dynamically relevant at intermediate redshifts.

The $\Lambda_{\rm s}$CDM model~\cite{Akarsu:2019hmw,Akarsu:2021fol,Akarsu:2022typ,Akarsu:2023mfb,Paraskevas:2024ytz,Yadav:2024duq,Akarsu:2024qsi,Akarsu:2024eoo,Akarsu:2024nas,Souza:2024qwd,Akarsu:2025gwi,Akarsu:2025dmj,Akarsu:2025ijk,Escamilla:2025imi,Akarsu:2025nns} (see also~\cite{Anchordoqui:2023woo,Anchordoqui:2024gfa,Anchordoqui:2024dqc,Soriano:2025gxd}) has attracted attention due to its ability to yield higher inferred values of $H_0$ and, in several analyses, improved Bayesian evidences compared to the standard $\Lambda$CDM scenario, while also alleviating other minor tensions such as the $S_8$ discrepancy~\cite{DiValentino:2020vvd}.\footnote{For further reading on theoretical and observational studies, as well as model-agnostic reconstructions, related to sign-changing dark-energy density in the late Universe, see the following (by no means exhaustive) list of references \cite{Sahni:2002dx,Vazquez:2012ag,BOSS:2014hwf,Sahni:2014ooa,BOSS:2014hhw,DiValentino:2017rcr,Mortsell:2018mfj,Poulin:2018zxs,Wang:2018fng,Banihashemi:2018oxo,Dutta:2018vmq,Banihashemi:2018has,Li:2019yem,Akarsu:2019ygx,Visinelli:2019qqu,Perez:2020cwa,Akarsu:2020yqa,Ruchika:2020avj,Calderon:2020hoc,DeFelice:2020cpt,Paliathanasis:2020sfe,Bonilla:2020wbn,Acquaviva:2021jov,Bag:2021cqm,Bernardo:2021cxi,Escamilla:2021uoj,Sen:2021wld,Ozulker:2022slu,DiGennaro:2022ykp,Akarsu:2022lhx,Moshafi:2022mva,vandeVenn:2022gvl,Ong:2022wrs,Tiwari:2023jle,Malekjani:2023ple,Vazquez:2023kyx,Alexandre:2023nmh,Adil:2023ara,Paraskevas:2023itu,Gomez-Valent:2023uof,Wen:2023wes,Wen:2024orc,DeFelice:2023bwq,Menci:2024rbq,Gomez-Valent:2024tdb,DESI:2024aqx,Bousis:2024rnb,Wang:2024hwd,Colgain:2024ksa,Tyagi:2024cqp,Toda:2024ncp,Sabogal:2024qxs,Dwivedi:2024okk,Escamilla:2024ahl,Gomez-Valent:2024ejh,Manoharan:2024thb,Pai:2024ydi,Mukherjee:2025myk,Efstratiou:2025xou,Gomez-Valent:2025mfl,Wang:2025dtk,Bouhmadi-Lopez:2025ggl,Tamayo:2025xci,Gonzalez-Fuentes:2025lei,Bouhmadi-Lopez:2025spo,Hogas:2025ahb,Yadav:2025vpx,Lehnert:2025izp,Tan:2025xas,Pedrotti:2025ccw,Forconi:2025gwo,Nyergesy:2025lyi,Ghafari:2025eql,Akarsu:2026anp}.} A defining feature of this model is that the effective energy density associated with the cosmological constant undergoes a sign change, becoming negative at earlier times. The phenomenological success of $\Lambda_{\rm s}$CDM has been attributed to this \emph{negative}
DE density phase, which can significantly affect the total cosmic energy budget at intermediate
redshifts, $z\sim2$, and lead to a considerable reduction in the Hubble rate at these redshifts. Within the CPL framework, a qualitatively similar effect is produced by a period of phantom-like behavior at comparable redshifts. Motivated by this analogy, we investigate whether introducing a negative-density phase into a more general DDE model can reduce the apparent need for NECB-crossing and bring the evolution closer to that of a constant equation of state.

We emphasize that, in the context of this work, crossing the PDL does not constitute a universal
classifier of dark-energy behavior. A central assumption of our analysis is that the effective
dark-energy density $\rho_{\rm DE}$ is \emph{not} required to be positive definite: within our
framework, $\rho_{\rm DE}$ is allowed to become negative and may even cross zero. Consequently,
the usual intuition based on the condition $w_{\rm DE}=-1$ cannot be used as a standalone
criterion to distinguish between ``quintessence-like'' and ``phantom-like'' regimes~\cite{Akarsu:2025gwi,Akarsu:2026anp}. Once sign-changing densities are admitted, the line
$w_{\rm DE}=-1$ ceases to be a reliable \emph{global} separator in the $(w,z)$ plane~\cite{Akarsu:2026anp}. The physically meaningful discriminator is instead the sign of
$\rho_{\rm DE}+p_{\rm DE}$, where $p_{\rm DE}$ denotes the dark-energy pressure, and the
corresponding boundary is the NECB, defined by $\rho_{\rm DE}+p_{\rm DE}=0$. This follows
directly from the continuity equation,
\begin{equation}
\frac{{\rm d}\rho_{\rm DE}}{{\rm d}\ln a}
= -3(\rho_{\rm DE}+p_{\rm DE})
= -3(1+w_{\rm DE})\,\rho_{\rm DE}\,,
\end{equation}
which shows that ``phantom-like'' evolution---in the sense that $\rho_{\rm DE}$ increases as
the Universe expands---corresponds to $\rho_{\rm DE}+p_{\rm DE}<0$, or equivalently
$(1+w_{\rm DE})\rho_{\rm DE}<0$. This condition implies
\begin{equation}
\rho_{\rm DE}+p_{\rm DE}<0
\quad \Longleftrightarrow \quad
\begin{cases}
w_{\rm DE}<-1, & \rho_{\rm DE}>0,\\[2pt]
w_{\rm DE}>-1, & \rho_{\rm DE}<0.
\end{cases}
\end{equation}
Therefore, when $\rho_{\rm DE}<0$, the correspondence between ``which side of $w_{\rm DE}=-1$''
and the actual physical regime is reversed. The resulting four kinematic branches---distinguished
by the signs of $\rho_{\rm DE}$ and $\rho_{\rm DE}+p_{\rm DE}$---are summarized in
\cref{tab:pq-branches}. Furthermore, if and when $\rho_{\rm DE}\rightarrow 0$ while $p_{\rm DE}$
remains finite, the equation-of-state parameter $w_{\rm DE}=p_{\rm DE}/\rho_{\rm DE}$ develops a
kinematic pole. Such a divergence does not signal a physical singularity in the background
evolution, but rather reflects the breakdown of $w_{\rm DE}$ as a diagnostic at a sign switch.
For this reason, to ensure clarity and internal consistency, we formulate our discussion
primarily in terms of $\rho_{\rm DE}$ and $\rho_{\rm DE}+p_{\rm DE}$ (or equivalently
$(1+w_{\rm DE})\rho_{\rm DE}$), and we avoid interpreting regime changes and NECB-crossing based
solely on the behavior of $w_{\rm DE}(z)$ without explicitly specifying the sign of
$\rho_{\rm DE}$~\cite{Carroll_2003,Adil:2023exv,Akarsu:2025gwi,Caldwell:2025inn,Akarsu:2026anp}.

The structure of the paper is as follows. In~\cref{sec:models}, we introduce the models considered in this work and describe how CPL-like evolution is combined with a negative dark energy density phase. In~\cref{sec:data_and_meth}, we outline the methodology adopted for cosmological inference and summarize the observational datasets employed. In~\cref{sec:results}, we present the results of our analysis, including parameter constraints, posterior distributions, and quantitative assessments of model performance. We also discuss the broader implications of our findings for future model building and for the interpretation of DDE scenarios. Finally, in~\cref{sec:conclusions}, we summarize the main conclusions of this study and place our results in a wider cosmological context.

\section{Models}
\label{sec:models}

In this work, we consider four cosmological models that introduce late-time departures from the standard $\Lambda$CDM scenario within the spatially flat Friedmann-Lema\^{i}tre-Robertson-Walker (FLRW) framework. We assume a cold dark matter (CDM) cosmology and explore four distinct phenomenological descriptions of DE.

For all models, the cosmic expansion as a function of the scale factor $a=(1+z)^{-1}$ is governed by the Friedmann equation
\begin{equation}
\label{eq:friedmann_general}
\frac{H^2(a)}{H_0^2}
=\frac{\Omega_{\rm r0}}{a^4}+\frac{\Omega_{\rm m0}}{a^3}
+\Omega_{\rm de0}\,\frac{\rho_{\rm DE}(a)}{\rho_0}\,,
\end{equation}
where $\Omega_{\rm de0}\equiv 1-\Omega_{\rm m0}-\Omega_{\rm r0}$ and $\rho_0\equiv \rho_{\rm DE}(a=1)$ denotes the present-day DE density. Here, $\Omega_{\rm r0}$, $\Omega_{\rm m0}$, and $\Omega_{\rm de0}$ represent the present-day density parameters of radiation, pressureless matter, and DE, respectively, and $H(a)$ is the Hubble parameter.
Any DE component characterized by an equation-of-state (EoS) parameter $w_{\rm DE}(a)$ satisfies the continuity equation
\begin{equation}
\label{eqn:cont_eqn}
\frac{{\rm d}\rho_{\rm DE}}{{\rm d}a} = -3\,\frac{\rho_{\rm DE}+p_{\rm DE}}{a}
= -3\,\frac{\rho_{\rm DE}\,(1+w_{\rm DE})}{a}\,,
\end{equation}
which leads to the general evolution of the energy density
\begin{equation}
\label{eqn:dens_evol_general}
|\rho_{\rm DE}(a)| = |\rho_{\rm DE}(a_*)|
\exp\!\left[-3\int_{\ln a_*}^{\ln a}\big[1+w_{\rm DE}(a')\big]\,{\rm d}\ln a' \right] \,,
\end{equation}
relative to a reference scale factor $a_*$. 
This equation holds piecewise on intervals where $\rho_{\rm DE}(a)\neq 0$ and therefore determines
the magnitude $|\rho_{\rm DE}|$; if $\rho_{\rm DE}$ is allowed to change sign (i.e.\ cross zero),
the sign must be specified separately by the chosen branch in phenomenological sign-switching
constructions.

Below, we detail how each of the four models are constructed differently based on these same principal equations. 

\subsection{Baseline CPL}
We consider the CPL scenario as our reference for model comparison. The CPL model is defined by its dynamical EoS
\begin{equation}
\label{eq:cpl_w}
w_{\rm CPL}(a)=w_0+(1-a)w_a,
\end{equation}
which implies the standard CPL density evolution
\begin{equation}
\label{eq:cpl_rho}
\rho_{\rm CPL}(a)=\rho_0\,a^{-3(1+w_0+w_a)}\,{\rm e}^{-3(1-a)w_a}.
\end{equation}
Here, $w_0$ and $w_a$ are constants that parametrize the linear EoS of CPL. The parameter $w_0 \equiv w_{\rm CPL}(a=1)$ represents the present-day value of the EoS, while $w_a$ quantifies the time dependence and controls the rate of deviation from $w_0$. 
The CPL form was originally introduced as a convenient reconstruction ansatz for a DE component whose behavior could be approximated by a Taylor expansion around its present-day value. The validity of this first-order truncation has been discussed and justified in Refs.~\cite{Planck:2015bue,Linder:2007ka}. Nevertheless, the CPL parametrization is not sufficiently general to describe several classes of DE models, such as those with singular EoS~\cite{Ozulker:2022slu} or scenarios involving negative energy densities~\cite{DiValentino:2020naf,Akarsu:2019hmw,Akarsu:2023mfb,Adil:2023exv,Specogna:2025guo}. Moreover, its linear form is not immune to possible reconstruction artifacts.

\subsection{CPL$_{>a_{\rm c}}$ (restricted-CPL control case)}

One potential reconstruction artifact associated with the linear form of $w_{\rm CPL}$ is the apparent
crossing of $w_{\rm DE}=-1$ reported in CPL-based reconstructions of the DESI DR2 results (see Fig.~12 of
Ref.~\cite{DESI:2025zgx}). In the standard positive-density setting, this is equivalent to reaching the
NECB, $\rho_{\rm DE}+p_{\rm DE}=0$.
Such NECB (often phrased as ``phantom-divide'') crossing is generally considered problematic because of the
difficulties in constructing consistent field-theory models that realize it~\cite{Cai:2009zp}, or because it
may point toward modified gravity or brane-world scenarios~\cite{Vagnozzi:2019ezj}. For these reasons, any apparent NECB-crossing signal should be interpreted with caution. 
Recently, two restricted variants of CPL were introduced in Ref.~\cite{Ozulker:2025ehg} that explicitly prohibit the DE EoS from exhibiting NECB-crossing by construction, with the aim of testing the robustness of the DESI results. In our analysis, we adopt their CPL$_{\rm >a_{\rm c}}$ model as a control case for comparison, for reasons that will become clear in~\cref{subsec:modelA}.

The CPL$_{\rm >a_{\rm c}}$ model employs the scale factor at which the CPL EoS in~\cref{eq:cpl_w} reaches the NECB to introduce a \textit{positive} cosmological constant phase in the DE EoS. The condition $w_{\rm CPL}(a_{\rm c})=-1$ defines the crossing scale factor $a_{\rm c}$ as
\begin{equation}
\label{eq:ac_def}
a_{\rm c} = 1 + \frac{1+w_0}{w_a}
\qquad (w_a \neq 0).
\end{equation}
The EoS is then described by the CPL form after the crossing and fixed to a $\Lambda$-like value before it:
\begin{equation}
\label{eq:modelA_w}
w_{\rm DE}(a)=
\begin{cases}
w_0+(1-a)w_a, & a>a_{\rm c},\\
-1,           & a\le a_{\rm c}.
\end{cases}
\end{equation}
Substituting this expression into~\cref{eqn:dens_evol_general} yields the corresponding energy density evolution
\begin{equation}
\label{eq:modelEmre_rho}
\rho_{\rm DE}(a)=
\begin{cases}
\rho_{\rm CPL}(a),    & a>a_{\rm c},\\
\rho_{\rm CPL}(a_{\rm c}),  & a\le a_{\rm c} ,
\end{cases}
\end{equation}
where continuity of $\rho_{\rm DE}$ at $a_{\rm c}$ is ensured by construction.

The authors of Ref.~\cite{Ozulker:2025ehg} concluded that NECB-crossing is indeed preferred, as evidenced by the significantly poorer fits of their restricted models compared to CPL in most cases. In contrast, in this work we investigate whether the phantom-like behavior favored by DESI BAO necessarily requires an evolving EoS that crosses the NECB. Alternative mechanisms capable of producing phantom DE could also reduce the expansion rate in the relevant redshift range and might provide an equally viable explanation. In particular, a negative DE energy density phase in the past can lead to a similar dynamical effect. Consequently, the apparent need for a dynamical EoS at $a>a_{\rm c}$ may be alleviated.

Before proceeding, we clarify what we mean by \emph{phantom} behavior. In the standard literature, DE with $w_{\rm DE}<-1$ is typically classified as phantom. Since most commonly studied DE models assume positive-definite energy densities, this identification is usually unambiguous. However, defining phantom behavior solely through a condition on the EoS becomes inadequate when DE is allowed to take non-positive values. 
More generally, phantom behavior corresponds to a DE density that decreases with decreasing redshift, that is, ${\rm d}\rho/{\rm d}a>0$. In this broader sense, when $\rho_{\rm DE}<0$, the corresponding EoS condition is inverted and phantom-like behavior is realized for $w_{\rm DE}>-1$. The theoretical foundations of this reframing, particularly from the perspective of scalar field theory, are discussed in Refs.~\cite{Akarsu:2025gwi,Akarsu:2026anp} (see also Ref.~\cite{Adil:2023exv}). 

In what follows, we introduce a modified version of CPL$_{\rm >a_{\rm c}}$ designed to test whether current data allow for a sign change in the effective DE density at the epoch where the CPL EoS would otherwise cross the NECB.

\subsection{CPL $\rightarrow -\Lambda$ (NECB-triggered sign-switch)}
\label{subsec:modelA}

Denoted as \textbf{CPL} $\bm{\rightarrow -\Lambda}$, this new model links the onset of a negative $\Lambda$ phase to the NECB-crossing of the CPL form. The EoS evolution of this model is identical to that of CPL$_{\rm >a_{\rm c}}$ in~\cref{eq:modelA_w}; however, its energy density undergoes a sign flip at $a_{\rm c}$:
\begin{equation}
\label{eq:modelA_rho}
\rho_{\rm DE}(a)=
\begin{cases}
\rho_{\rm CPL}(a),    & a>a_{\rm c},\\
-\rho_{\rm CPL}(a_{\rm c}), & a\leq a_{\rm c}.
\end{cases}
\end{equation}
This abrupt-step sign switch is adopted here as a controlled phenomenological analogue of abrupt
implementations of sign-changing vacuum energy, as considered in the abrupt $\Lambda_{\rm s}$CDM
literature (see, e.g., Refs.~\cite{Akarsu:2019hmw,Akarsu:2021fol,Akarsu:2022typ,Akarsu:2023mfb}). The switching epoch is fixed by the
CPL-inferred NECB-crossing scale factor $a_{\rm c}$, while the post-transition evolution is governed
by the CPL branch. This construction realizes an effective negative cosmological constant at early times, while matching the CPL branch in magnitude at the transition. In practice, the model behaves as a CPL DDE at late times and approaches a constant \emph{negative} cosmological constant energy density in the pre-transition era.
A limitation of this piecewise prescription is that $\rho_{\rm DE}$ is formally discontinuous at $a=a_{\rm c}$. A fully predictive theoretical realization would instead feature a continuous sign switch occurring
over a finite interval in $a$ (or $z$), as discussed, for example, in
Refs.~\cite{Akarsu:2024qsi,Akarsu:2025gwi}.
For the purposes of this phenomenological analysis, we adopt this step-like idealization as an
effective limit of a sufficiently rapid transition: the likelihoods used here are dominated by
background distance observables, which depend on $H(z)$ and on redshift integrals thereof, and are
therefore expected to be largely insensitive to the value of $\rho_{\rm DE}$ at an isolated switching
point (and, in practice, are further smoothed by the finite redshift resolution of the data).
Additional motivation for the abrupt limit comes from the abrupt $\Lambda_{\rm s}$CDM analysis of
Ref.~\cite{Paraskevas:2024ytz}, where implementing a late-time ($z\sim2$) AdS-to-dS sign switch as a
step-like transition was reported to have only a weak impact on the evolution of bound cosmic
structures.
Our primary objective is to assess whether the apparent preference for NECB crossing persists,
weakens, or changes character once a negative DE density phase is admitted.

Although Ref.~\cite{Ozulker:2025ehg} introduced an additional variant in which the (positive) cosmological constant phase occurs after the CPL era, we do not consider that construction here. A negative cosmological constant at the present epoch would lead to a recollapsing universe on timescales incompatible with the observed cosmic age. For this reason, we adopt only their CPL$_{>a_{\rm c}}$ model as a control case and focus instead on the modification outlined above, where the positive cosmological constant phase is replaced by a negative energy density phase.

Both CPL$\rightarrow -\Lambda$ and CPL$_{>a_{\rm c}}$ contain CPL-only and $\mp\Lambda$-only regimes when $a_{\rm c}$ lies outside the physical interval $[0,1]$. If the formal crossing occurs at $a_{\rm c}<0$, the transition takes place before the onset of the standard cosmological evolution, and the model effectively reduces to the standard CPL scenario with no realized transition. Conversely, if $a_{\rm c}>1$, the transition is pushed into the future and the model remains entirely in the $\mp\Lambda$ phase throughout the observable history. 
Importantly, these non-transitioning regimes cannot be interpreted simply as limits of the free parameters, since $a_{\rm c}$ is a nonlinear function of $(w_0,w_a)$. Consequently, exploring parameter regions with $a_{\rm c}\notin[0,1]$ complicates the interpretation of model performance and the associated parameter constraints.

To avoid the complications discussed above, we restrict the analysis to scenarios with $a_{\rm c} \in [0,1]$ and exclude the corresponding $(w_0,w_a)$ combinations that place the transition outside this interval. This procedure necessarily reduces the effective two-dimensional $w_0$–$w_a$ parameter space relative to that of $w_0w_a$CDM. However, this restriction does not compromise our objectives, since the purpose of these constructions is specifically to test whether the observed NECB-crossing and dynamical behavior of CPL-parametrized DE represent genuine physical features rather than reconstruction artifacts.

It is also important to emphasize that both CPL$\rightarrow -\Lambda$ and CPL$_{>a_{\rm c}}$ cannot simultaneously reproduce an exact cosmological constant behavior at $a>a_{\rm c}$ and realize their respective post-transition phases. In particular, since $a_{\rm c} \rightarrow \infty$ when $w_a\to0$, these models cannot continuously approach the $\Lambda$ limit while retaining a transition within the physical interval. As a result, they must balance between adopting $w_a \simeq 0$ at $a>a_{\rm c}$ and exhibiting a (negative or positive) $\Lambda$ phase at $a<a_{\rm c}$. 
This limitation is especially relevant for the CPL$\rightarrow -\Lambda$ construction, which by design cannot smoothly reduce to the $\Lambda_{\rm s}$CDM framework~\cite{Akarsu:2023mfb,Akarsu:2024qsi}. For this reason, we introduce the sCPL model in the next section, which allows for that limit.

\subsection{sCPL (free sign-switch in $\rho_{\rm DE}$)}
\label{subsec:modelB}

Denoted as \textbf{sCPL}, this new model retains the CPL EoS at all times,
\begin{equation}
\label{eq:modelB_w}
w_{\rm DE}(a)=w_0+(1-a)w_a
\qquad \text{for all } a,
\end{equation}
but introduces an \emph{independent} transition scale factor $a_\dagger$
(equivalently $z_\dagger=a_\dagger^{-1}-1$), at which the DE density changes sign:
\begin{equation}
\label{eq:modelB_rho}
\rho_{\rm DE}(a)=
\begin{cases}
\rho_{\rm CPL}(a),   & a>a_\dagger,\\
-\rho_{\rm CPL}(a),  & a\leq a_\dagger,
\end{cases}
\end{equation}
an abrupt-step idealization reminiscent of the abrupt $\Lambda_{\rm s}$CDM
scenario~\cite{Akarsu:2021fol}, which may be written schematically as
$\Lambda_{\rm s}(a)=\Lambda_{\rm s0}\,\mathrm{sgn}(a-a_\dagger)$, where
$\Lambda_{\rm s0}$ denotes the post-transition (i.e.\ present-day) value.
In the limit $w_0\rightarrow -1$ and $w_a\rightarrow 0$, this construction
reduces to the abrupt $\Lambda_{\rm s}$CDM framework~\cite{Akarsu:2019hmw,Akarsu:2021fol,Akarsu:2022typ,Akarsu:2023mfb}.
As in the CPL$\rightarrow -\Lambda$ case, the sign flip at $a=a_\dagger$
formally introduces a discontinuity in $\rho_{\rm DE}$.
A fully predictive realization would instead implement the sign switch
continuously over a finite interval in $a$ (or $z$); here we adopt the same
sharp-step idealization and justification discussed in~\cref{subsec:modelA},
to be understood as the effective limit of a sufficiently rapid transition.

\begin{figure}[h!tb]
    \centering
    \includegraphics[width=1. \columnwidth]{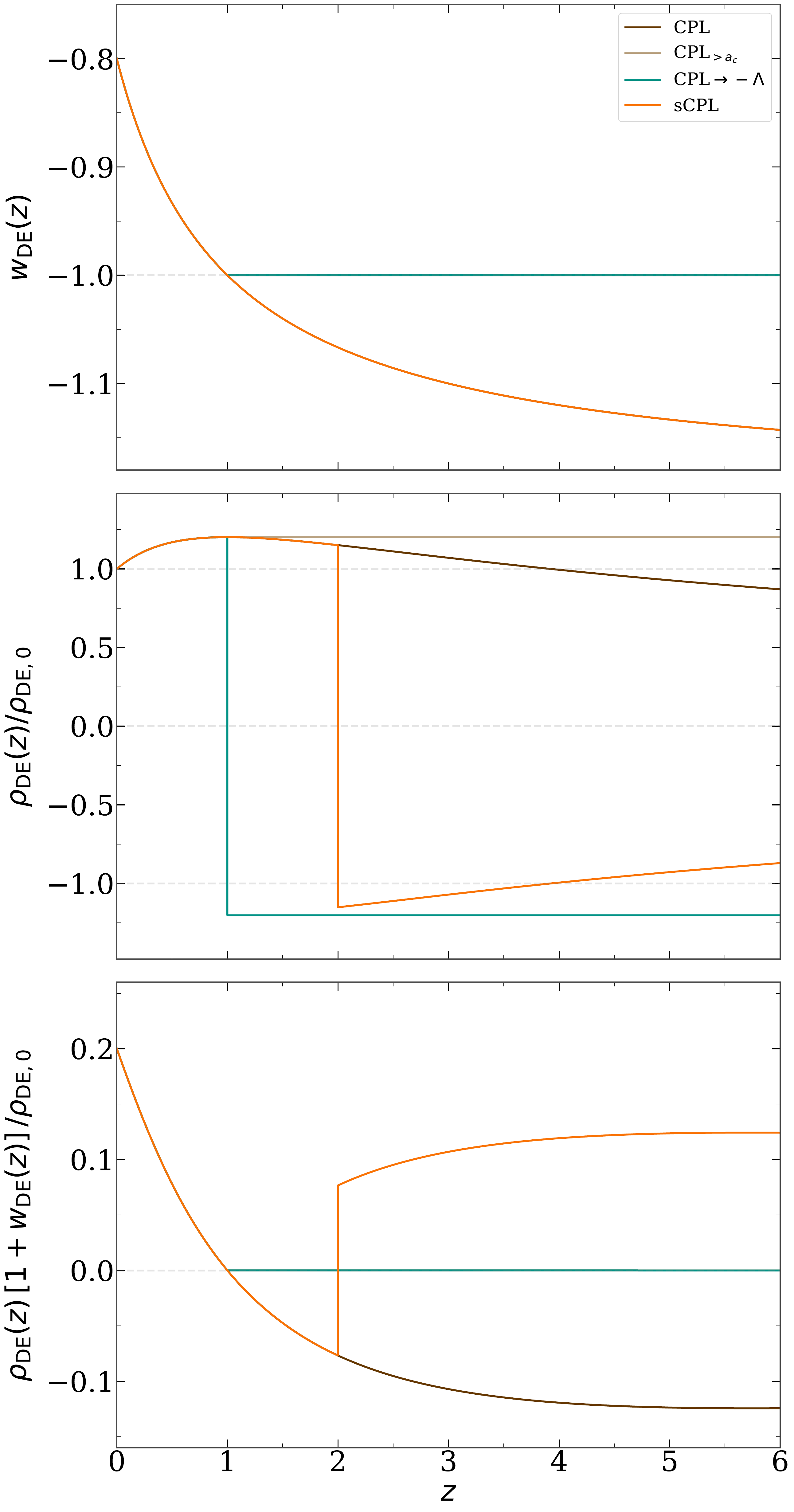}
    \caption{Evolution of the DE EoS (\textbf{top panel}), energy density (\textbf{middle panel}), 
and $\rho_{\rm DE}+p_{\rm DE}$ (\textbf{bottom panel}) for the models considered, assuming 
$w_0=-0.8$, $w_a=-0.4$, and $z_\dagger=2$. In the top panel, the CPL and CPL$_{>a_{\rm c}}$ curves are 
indistinguishable from those of sCPL and CPL$\to -\Lambda$, respectively, and are therefore not 
separately visible. Likewise, in the bottom panel, CPL$_{>a_{\rm c}}$ and CPL$\to -\Lambda$ exhibit 
identical evolutions.}
    \label{fig:models_evol}
\end{figure}

\cref{fig:models_evol} illustrates the distinct behaviors of the proposed models for identical parameter choices. Compared to CPL$\to -\Lambda$, the key conceptual difference is that the switching epoch in sCPL is treated as an independent free parameter, $a_\dagger$ (or equivalently $z_\dagger$), and is not tied to the NECB-crossing of the CPL EoS. We impose priors such that the transition occurs in the past, thereby avoiding a negative DE density today, and that it lies approximately within the redshift range probed by late-time observations, as detailed in~\cref{sec:data_and_meth}. If the transition is pushed sufficiently early (i.e.\ $a_\dagger\ll 1$, or equivalently $z_\dagger$ is large),
the model effectively reduces to CPL behavior over the redshift range probed by the data, rendering the
analysis insensitive to the underlying sign-switch mechanism.

\begin{table}[t!]
    \centering
    \setlength{\tabcolsep}{10pt}
    \renewcommand{\arraystretch}{1.0}
    \begin{tabular}{cccc}
        \toprule
         $\rho_{\text{DE}}$ & $w_{\text{DE}}$ & $\rho_{\text{DE}} \, (1+ w_{\text{DE}})$ & Classification \\
        \midrule
        $>0$ & $>-1$ & $>0$ & $p$-quintessence \\
        $<0$ & $<-1$ & $>0$ & $n$-quintessence \\
        \addlinespace
        $>0$ & $<-1$ & $<0$ & $p$-phantom \\
        $<0$ & $>-1$ & $<0$ & $n$-phantom \\
        \bottomrule
    \end{tabular}
    \caption{Classification of phantom and quintessence branches from the perspective of a single scalar field theory, see, e.g., Refs.~\cite{Akarsu:2025gwi,Akarsu:2026anp} (see also Ref.~\cite{Adil:2023exv}). The prefixes $p$ and $n$ denote $\rho_{\rm DE}>0$ and $\rho_{\rm DE}<0$, respectively.}
    \label{tab:pq-branches}
\end{table} 

Within the framework of our clarified definition of phantom DE, the sCPL model can exhibit a nontrivial history characterized by alternating quintessence (q) and phantom (p) regimes. Since the sign-switch is decoupled from the NECB-crossing, the ordering of events can lead to either $p \to q \to p$ or $q \to p \to q$ behavior, where the arrows denote increasing redshift. We further subdivide the phantom and quintessence regimes according to the sign of $\rho_{\rm DE}$ and the position of $w_{\rm DE}$ relative to the NECB, as summarized in~\cref{tab:pq-branches}. 
For example, if $w_0>-1$, $w_a<0$, and $z_{\rm c}<z_\dagger$, the model follows a $p$-quintessence $\to$ $p$-phantom $\to$ $n$-quintessence sequence. Conversely, if $w_0<-1$, $w_a>0$, and $z_{\rm c}<z_\dagger$, the evolution becomes $p$-phantom $\to$ $p$-quintessence $\to$ $n$-phantom. This possible alternation of regimes is illustrated in the bottom panel of \cref{fig:models_evol}. 
For the representative parameter choices shown, the sCPL model exhibits a sequence of 
$p$-quintessence at $z\simeq 0$--$1$, followed by $p$-phantom behavior at $z\simeq 1$--$2$, 
and finally $n$-quintessence at $z\gtrsim 2$. 
This example demonstrates that the equation-of-state parameter $w_{\rm DE}$ alone is insufficient 
to characterize phantom or quintessence behavior, and highlights the fundamentally misleading nature 
of interpreting the condition $w_{\rm DE}=-1$ as a universal ``phantom divide line''.

\section{Data and Methodology} \label{sec:data_and_meth}

We perform cosmological inference for the aforementioned models using the Monte Carlo Markov Chain (MCMC) sampling method implemented within the \texttt{Cobaya} framework~\cite{Torrado:2020dgo}. Theoretical predictions are computed with the \texttt{CAMB} code~\cite{Lewis:1999bs,Howlett_2012}, where DE perturbations are treated using the parametrized post-Friedmann (PPF) prescription~\cite{Fang:2008sn}. Convergence of the MCMC chains is assessed using the Gelman--Rubin criterion~\cite{Gelman:1992zz}, requiring $R-1<0.01$. The resulting chains are analyzed and visualized with the \texttt{GetDist} package~\cite{Lewis:2019xzd}.

In addition to the six free parameters of the $\Lambda$CDM model, $w_0w_a$CDM introduces two extra parameters, $w_0$ and $w_a$, to describe a dynamical DE EoS. CPL$\rightarrow -\Lambda$ model is characterized by the same set of free parameters as $w_0w_a$CDM. sCPL model, however, extends the parameter space by introducing the additional transition parameter $z_\dagger$, allowing for an independent sign-switch epoch. For all models, we adopt flat priors over the parameter ranges specified in~\cref{tab:priors}.

\begin{table}[hbt]
    \caption{Model parameters and the corresponding flat prior ranges used in the cosmological inference.} 
    \centering
    \begin{tabular}{c c}
        \toprule
         \, Parameter\, & Prior \\
         \midrule
         $\Omega_{\rm b} h^2$ & $[0.005, 0.1]$ \\
         $\Omega_{\rm c} h^2$ & $[0.001, 0.99]$ \\
         $\tau_{\rm reio}$ & $[0.01, 0.8]$ \\
         $\ln(10^{10}A_s)$ & $[1.61, 3.91]$ \\
         $n_s$ & $[0.8, 1.2]$ \\
         $\theta_s$ & $[0.5, 10]$ \\
         \midrule
         $w_0$ & $[-3, 1]$ \\
         $w_a$ & $[-3, 2]$ \\
         \midrule
         $z_\dagger$ & $[1, 3]$ \\
         \bottomrule
    \end{tabular}
    \label{tab:priors}
\end{table}

For the observational datasets, we incorporate both early- and late-time measurements in various combinations.

For CMB anisotropies and lensing, we use the \texttt{Planck 2018}~\cite{Planck:2019nip,Planck:2018lbu} power spectra. Specifically, we include the low-$\ell$ ($\ell<30$) $TT$ \texttt{Commander} likelihood, the low-$\ell$ $EE$ \texttt{SimAll} likelihood, and the high-$\ell$ ($\ell\geq 30$) \texttt{CamSpec} likelihood~\cite{Efstathiou:2019mdh,Rosenberg:2022sdy}, together with the \texttt{NPIPE PR4} lensing likelihood~\cite{Carron:2022eyg}.

BAO distance measurements from \texttt{DESI DR2}~\cite{DESI:2025zgx} play a central role in our analysis. Covering an effective redshift range of $0.295<z<2.330$, the DESI BAO data probe the expansion history and large-scale structure, where the dark sector contributes significantly~\cite{Weinberg:2013agg}.

We also include three independent SNeIa compilations to assess how the models describe the recent expansion history. The \texttt{Pantheon+} sample~\cite{Scolnic:2021amr} contains 1550 objects in the range $0.001<z<2.26$. The Dark Energy Survey Year 5 (\texttt{DESY5}) dataset~\cite{DES:2024jxu} comprises 1635 SNeIa from a single survey spanning $0.10<z<1.13$; we adopt the latest \texttt{Dovekie} recalibration~\cite{Popovic:2025glk,DES:2025sig}. Finally, the \texttt{Union3} compilation~\cite{Rubin:2023jdq} includes 2087 objects. Although these datasets share a substantial number of supernovae, they remain statistically distinct due to differences in calibration strategies, systematic treatments, and light-curve fitting methodologies. 

After the data analysis is complete, we calculate the Akaike information criterion (AIC)~\cite{1100705,Liddle:2007fy}
\begin{equation}
    \label{eqn:aic}
    \text{AIC}_{\rm model} \equiv -2\ln \mathcal{L}_{\rm max} +2k 
    = \chi^2_{\rm min} +2k
\end{equation}
for each model and dataset combination as a measure of relative goodness of fit, where $\mathcal{L}_{\rm max}$ is the maximum likelihood achieved by the model, $k$ is the number of free parameters, and $\chi^2_{\rm min}$ is the minimum chi-square value. By explicitly accounting for $k$, the AIC allows for a fair comparison between models with different numbers of parameters, as is the case for sCPL relative to the other constructions. A positive $\Delta$AIC $\equiv {\rm AIC}_{\rm model} - {\rm AIC}_{\rm CPL}$ indicates a worse fit to the data with respect to the reference CPL model.

In addition to the AIC, we compute the Bayesian evidence
\begin{equation}
    \mathcal{Z}_i =\mathcal{Z}(d \, | \, \mathcal{M}_i ) \equiv \int \,d \theta \ p(d \, |\, \theta,\mathcal{M}_i) \ p(\theta \, | \, \mathcal{M}_i)
\end{equation}
using a modified version of the \texttt{MCEvidence}\footnote{\href{https://github.com/williamgiare/wgcosmo/tree/main}{github.com/williamgiare/wgcosmo/tree/main}}$^,$\footnote{\href{https://github.com/yabebalFantaye/MCEvidence}{github.com/yabebalFantaye/MCEvidence}} package~\cite{Heavens:2017hkr,Heavens:2017afc}, fully compatible with \texttt{Cobaya}. Here, $\mathcal{L}= p(d \, |\, \theta,\mathcal{M}_i)$ denotes the likelihood and $p(\theta \, | \, \mathcal{M}_i)$ the prior distribution. The quantities $d$, $\theta$, and $\mathcal{M}_i$ represent the observational data, the model parameters, and a specific model, respectively. The Bayes factor $\mathcal{B}_{ij}$ is then defined as
\begin{equation}\label{eqn:bayesfac}
    \ln\mathcal{B}_{ij} \equiv \ln \mathcal{Z}_i -  \ln \mathcal{Z}_j
\end{equation}
with respect to the reference CPL model, denoted by the subscript $j$. We interpret $\mathcal{B}_{ij}$ using the Jeffreys scale, as summarized in Tab.~1 of Ref.~\cite{Trotta:2008qt}. A Bayes factor satisfying $|\ln\mathcal{B}_{ij}| <1$ corresponds to inconclusive evidence, meaning that the two models perform comparably. If $1 \lesssim |\ln\mathcal{B}_{ij}| \lesssim 2.5$, $2.5 \lesssim |\ln\mathcal{B}_{ij}| \lesssim 5$, or $|\ln\mathcal{B}_{ij}| \gtrsim 5$, the evidence against the model under consideration is weak, moderate, or strong, respectively. In our convention, negative values of $\ln\mathcal{B}_{ij}$ indicate poorer performance of $\mathcal{M}_i$ relative to CPL. 
Although the Bayesian evidence provides a more comprehensive assessment of model performance by accounting for prior volume effects and the full posterior structure, compared to $\chi^2$-based criteria, the AIC remains a useful indicator of relative goodness of fit. For this reason, we consider both $\Delta$AIC and $\mathcal{B}_{ij}$ in our model comparison.

\section{Results}
\label{sec:results}

\begin{table*}[p]
\caption{Constraints for the models considered. We report the best-fit $\Delta{\rm AIC}\equiv {\rm AIC}_{\rm model}-{\rm AIC}_{\rm CPL}$ and the Bayes factors $\mathcal{B}_{ij}$, defined as $\ln B_{ij} = \ln \mathcal{Z}_{\rm model} - \ln \mathcal{Z}_{\rm CPL}$, where CPL is the reference model. Positive $\Delta{\rm AIC}$ values indicate worse fits to the data, whereas negative values of $\ln\mathcal{B}_{ij}$ indicate poorer performance of model $\mathcal{M}_i$ relative to CPL. The quantity $z_{\rm c}$ denotes the (derived) redshift of NECB-crossing, i.e. the redshift corresponding to $a_{\rm c}$ when such a crossing occurs within the relevant domain. For $z_{\rm c}$ entries marked by *, we report the upper and lower 68\% confidence bounds around the posterior peak (rather than the mean) when the mean lies outside the credible interval, which can occur for distributions with long tails. For one-sided constraints, we quote the 95\% confidence interval.}
\label{tab:constraints_main}
\centering
\footnotesize
\setlength{\tabcolsep}{3pt} 
\renewcommand{\arraystretch}{1.15}

\resizebox{\textwidth}{!}{%
\begin{tblr}{
  colspec = {l *{7}{c} r},
  row{1}  = {font=\bfseries},
  rowsep  = 2pt,
  colsep  = 5pt,
}
\toprule
Model / Dataset
& $w_0$ & $w_a$ & $z_\dagger$& $\Omega_{\rm m0}$ & $H_0 \, [\rm km/s/ Mpc]$  & $z_{\rm c}$ & $\Delta {\rm AIC}$ & $\ln \mathcal{B}_{ij}$ \\
\midrule

\SetCell[c=9]{l}{\bfseries \boldmath CPL $\rightarrow -\Lambda$ } \\

Planck
& $-1.83^{+0.14}_{-0.40}$ & $1.12^{+0.58}_{-0.33}$ & \textemdash
& $0.1864^{+0.0070}_{-0.044}$ & $> 73.4$ & $2.04_{-0.31}^{+6.22}$ * & $0.55$ & $-2.19$ \\

Planck+Pantheon+
& $-0.904^{+0.040}_{-0.055}$ & $-0.138^{+0.097}_{-0.056}$ & \textemdash
& $0.309^{+0.020}_{-0.012}$ & $67.9^{+1.1}_{-2.1}$ & $2.02_{-0.22}^{+3.80}$ * & $1.25$ & $-3.28$ \\

Planck+DESY5
& $-0.882^{+0.048}_{-0.067}$ & $-0.177^{+0.13}_{-0.070}$ & \textemdash
& $0.304^{+0.022}_{-0.014}$ & $68.5^{+1.3}_{-2.4}$ & $1.83_{-0.19}^{+2.37}$ * & $2.30$ & $-4.10$ \\

Planck+Union3
& $-0.820^{+0.062}_{-0.081}$ & $-0.257^{+0.15}_{-0.088}$ & \textemdash
& $0.322^{+0.022}_{-0.016}$ & $66.6^{+1.4}_{-2.3}$ & $1.90_{-0.14}^{+3.59}$ * & $2.71$ & $-3.99$ \\

Base: Planck+DESI
& $-0.99\pm 0.11$ & $-0.02\pm 0.15$ & \textemdash
& $0.303\pm 0.016$ & $68.5\pm 1.8$ & $2.67_{-0.15}^{+1.34}$ * & $5.51$ & $-6.19$ \\

Base+Pantheon+
& $-0.925\pm 0.033$ & $-0.102^{+0.046}_{-0.040}$ & \textemdash
& $0.3119\pm 0.0054$ & $67.43\pm 0.56$ & $2.61_{-0.12}^{+0.76}$ * & $1.15$ & $-4.29$ \\

Base+DESY5
& $-0.921^{+0.028}_{-0.032}$ & $-0.107^{+0.045}_{-0.038}$ & \textemdash
& $0.3124\pm 0.0050$ & $67.38\pm 0.52$ & $2.57_{-0.08}^{+0.76}$ * & $2.87$ & $-5.20$ \\

Base+Union3
& $-0.881\pm 0.045$ & $-0.163^{+0.065}_{-0.058}$ & \textemdash
& $0.3184\pm 0.0070$ & $66.75\pm 0.73$ & $2.58_{-0.12}^{+0.59}$ * & $5.46$ & $-6.51$ \\

\midrule
\SetCell[c=9]{l}{\bfseries sCPL} \\

Planck
& $-1.19^{+0.44}_{-0.61}$ & $< 1.29$ & ---
& $0.220^{+0.019}_{-0.078}$ & $>63.3$ & $-0.033_{-0.564}^{+0.694}$ * & $3.36$ & $-0.30$ \\

Planck+Pantheon+
& $-0.859\pm0.090$ & $-0.48\pm0.49$ & $>1.70$
& $0.302^{+0.011}_{-0.010}$ & $68.6^{+1.0}_{-1.3}$ & $0.330_{-0.329}^{+0.244}$ * & $3.46$ & $-0.80$ \\

Planck+DESY5
& $-0.772\pm0.099$ & $-0.89\pm0.51$ & $>1.58$
& $0.302^{+0.010}_{-0.0084}$ & $68.66^{+0.81}_{-1.1}$ & $0.302^{+0.12}_{-0.060}$ & $4.29$ & $-0.75$ \\

Planck+Union3
& $-0.64\pm0.14$ & $-1.30\pm0.67$ & $>1.58$
& $0.313\pm0.013$ & $67.5\pm1.3$ & $0.379^{+0.095}_{-0.11}$ & $2.23$ & $-0.76$ \\

Base: Planck+DESI
& $-0.50\pm0.23$ & $-1.45\pm0.67$ & $>2.37$
& $0.346\pm0.023$ & $64.3^{+1.9}_{-2.3}$ & $0.5\pm4.3$ & $3.82$ & $-1.13$ \\

Base+Pantheon+
& $-0.870\pm0.058$ & $-0.42\pm0.23$ & $>2.38$
& $0.3112\pm0.0057$ & $67.68\pm0.60$ & $0.45^{+0.22}_{-0.14}$ & $2.89$ & $-0.63$ \\

Base+DESY5
& $-0.837\pm0.060$ & $-0.55\pm0.25$ & $>2.38$
& $0.3128\pm0.0053$ & $67.52\pm0.54$ & $0.359^{+0.15}_{-0.062}$ & $2.40$ & $-0.93$ \\

Base+Union3
& $-0.704\pm0.093$ & $-0.91\pm0.32$ & $>2.38$
& $0.3261\pm0.0088$ & $66.14\pm0.87$ & $0.545^{+0.063}_{-0.14}$ & $2.20$ & $-1.13$ \\

\midrule
\SetCell[c=9]{l}{\bfseries \boldmath CPL$_{>a_{\rm c}}$ (control)} \\

Planck & $-1.17^{+0.77}_{-1.0}$ & \textemdash & \textemdash
& $0.286^{+0.078}_{-0.14}$ & $73^{+20}_{-20}$ & $< 7.22$ & $0.97$ & $-0.58$ \\

Planck+Pantheon+ & $-0.864^{+0.087}_{-0.11}$ & $< 0.392$ & \textemdash
& $0.3208\pm 0.0069$ & $66.69\pm 0.57$ & $< 0.273$ & $0.13$ & $2.17$ \\

Planck+DESY5 & $-0.792^{+0.12}_{-0.091}$ & $< -0.456$ & \textemdash
& $0.3236\pm 0.0069$ & $66.37\pm 0.59$ & $0.149^{+0.050}_{-0.10}$ & $1.31$ & $1.59$ \\

Planck+Union3 & $-0.67^{+0.16}_{-0.12}$ & $< -0.455$ & \textemdash
& $0.336\pm 0.011$ & $65.19^{+0.94}_{-1.1}$ & $0.282^{+0.044}_{-0.16}$ & $0.57$ & $0.76$ \\

Base: Planck+DESI & $-0.79\pm 0.23$ & $< 0.939$ & \textemdash
& $0.314^{+0.013}_{-0.017}$ & $67.0^{+1.7}_{-1.5}$ & $< 0.517$ & $10.96$ & $-2.59$ \\

Base+Pantheon+ & $-0.821^{+0.11}_{-0.094}$ & $< -0.272$ & \textemdash
& $0.3097^{+0.0052}_{-0.0060}$ & $67.39^{+0.61}_{-0.52}$ & $0.159^{+0.026}_{-0.12}$ & $7.12$ & $-0.28$ \\

Base+DESY5 & $-0.738^{+0.11}_{-0.080}$ & $< -0.559$ & \textemdash
& $0.3140\pm 0.0057$ & $66.90^{+0.53}_{-0.60}$ & $0.191^{+0.041}_{-0.094}$ & $5.62$ & $-0.37$ \\

Base+Union3 & $-0.63^{+0.15}_{-0.11}$ & $< -0.625$ & \textemdash
& $0.3246\pm 0.0094$ & $65.80^{+0.83}_{-1.0}$ & $0.272^{+0.069}_{-0.12}$ & $8.50$ & $-2.06$ \\

\midrule
\SetCell[c=9]{l}{\bfseries CPL (reference)} \\

Planck & $-1.16^{+0.49}_{-0.67}$ & --- & \textemdash & $0.237^{+0.025}_{-0.094}$ & $80^{+20}_{-7}$ & $-0.11^{+0.88}_{-0.63}$ & $0.00$ & $0.00$ \\
Planck+Pantheon+ & $-0.878\pm0.097$ & $-0.51^{+0.51}_{-0.43}$ & \textemdash & $0.313^{+0.011}_{-0.013}$ & $67.5\pm1.2$ & $0.227^{+0.24}_{-0.096}$ & $0.00$ & $0.00$ \\
Planck+DESY5 & $-0.79\pm0.10$ & $-0.90\pm0.50$ & \textemdash & $0.3100^{+0.0087}_{-0.0099}$ & $67.78\pm0.94$ & $0.377^{+0.020}_{-0.122}$ & $0.00$ & $0.00$ \\
Planck+Union3 & $-0.66\pm0.14$ & $-1.28\pm0.66$ & \textemdash & $0.320^{+0.012}_{-0.014}$ & $66.7\pm1.3$ & $0.395_{-0.098}^{+0.108}$ * & $0.00$ & $0.00$ \\
Base: Planck+DESI & $-0.44\pm0.21$ & $-1.66\pm0.59$ & \textemdash & $0.351\pm0.021$ & $63.8^{+1.7}_{-2.1}$ & $0.500^{+0.069}_{-0.054}$ & $0.00$ & $0.00$ \\
Base+Pantheon+ & $-0.842\pm0.054$ & $-0.59^{+0.21}_{-0.19}$ & \textemdash & $0.3111\pm0.0057$ & $67.50\pm0.59$ & $0.38\pm0.53$ & $0.00$ & $0.00$ \\
Base+DESY5 & $-0.812\pm0.056$ & $-0.69^{+0.23}_{-0.20}$ & \textemdash & $0.3131\pm0.0053$ & $67.32\pm0.55$ & $0.385^{+0.052}_{-0.085}$ & $0.00$ & $0.00$ \\
Base+Union3 & $-0.674\pm0.088$ & $-1.05^{+0.31}_{-0.27}$ & \textemdash & $0.3270\pm0.0088$ & $65.91\pm0.85$ & $0.462^{+0.055}_{-0.081}$ & $0.00$ & $0.00$ \\

\bottomrule
\end{tblr}
}
\end{table*}

\begin{figure*}[h!tb]
    \centering
    \includegraphics[width=0.95\columnwidth]{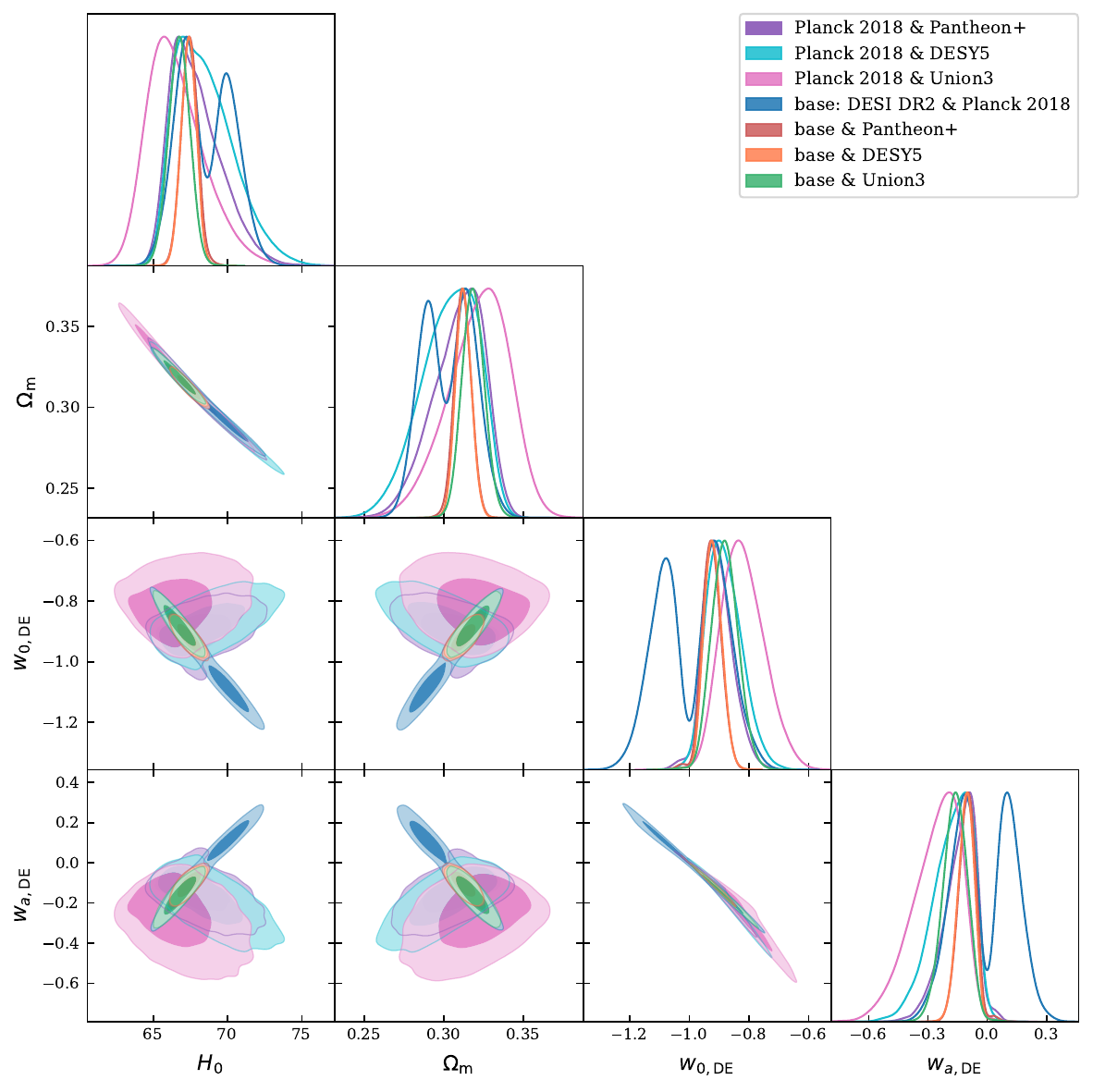}\hfill
    \includegraphics[width=0.95\columnwidth]{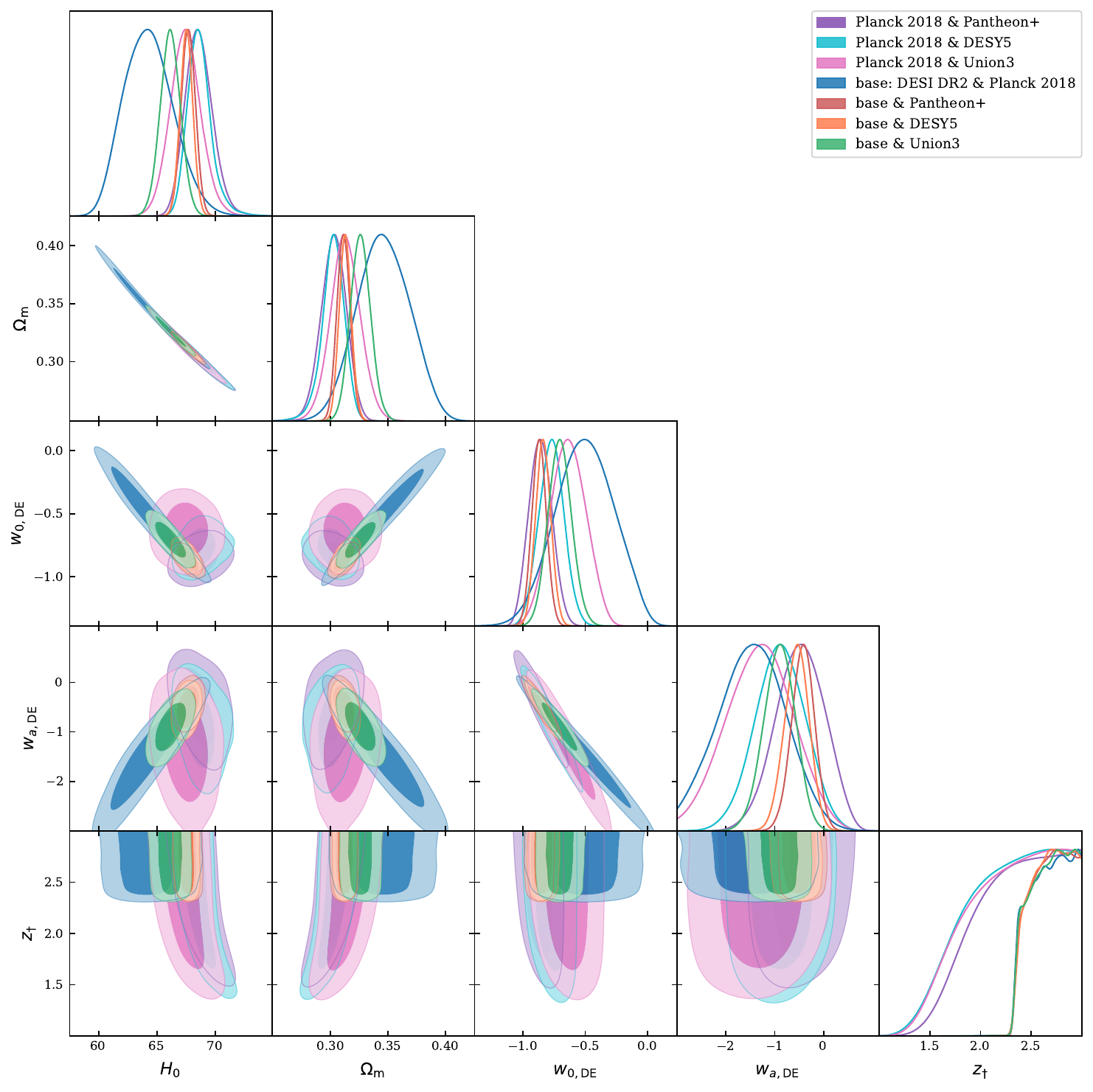}\hfil\\
    [-2pt]
    \parbox[t]{0.95\columnwidth}{\centering (a) CPL $\rightarrow -\Lambda$}\hfill
    \parbox[t]{0.95\columnwidth}{\centering (b) sCPL} \hfil\\
    [10pt]
    \includegraphics[width=0.95\columnwidth]{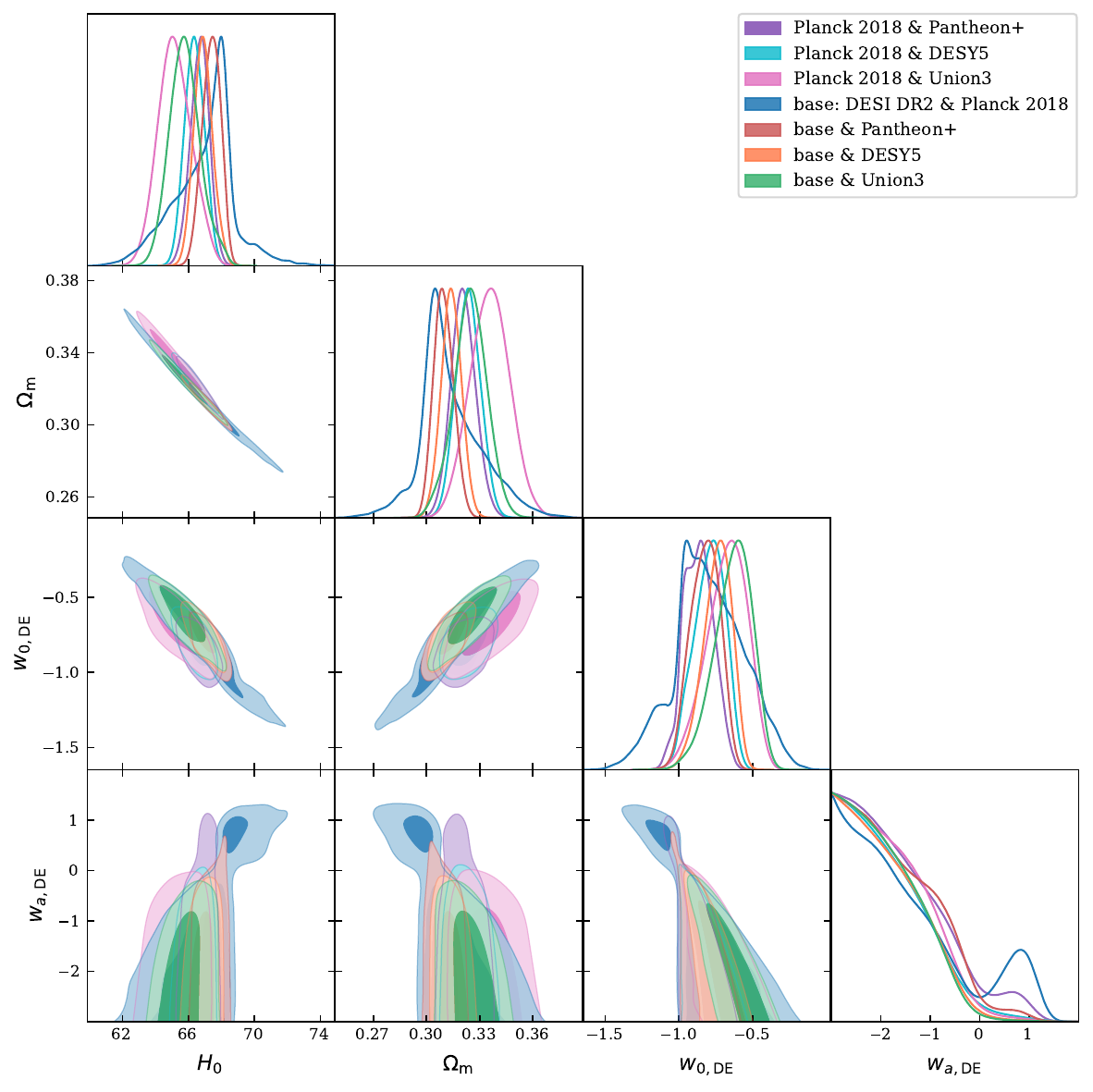}\hfil
    \includegraphics[width=0.95\columnwidth]{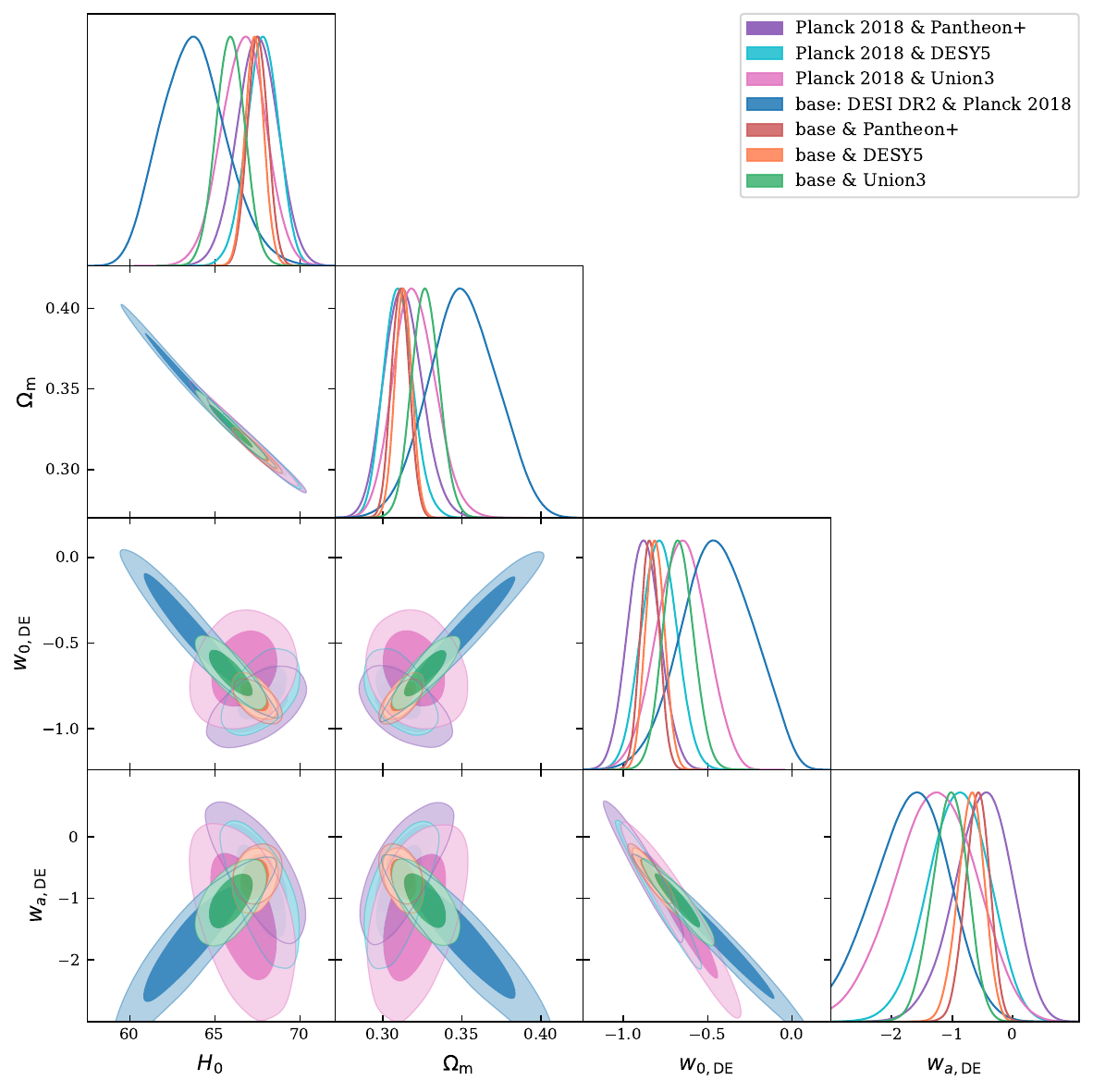}\hfil\\
    [-2pt]
    \parbox[t]{0.95\columnwidth}{\centering (c) CPL$_{>a_{\rm c}}$} \hfill
    \parbox[t]{0.95\columnwidth}{\centering (d) CPL} \hfil\\
    
    \caption{One- and two-dimensional marginalized posterior distributions of the model parameters for the four models considered, obtained using various dataset combinations.}
    \label{fig:posteriors_permodel}
\end{figure*}

\begin{figure*}[th!]
    \centering
    \includegraphics[width=0.19\linewidth]{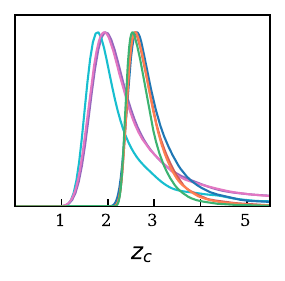} \hfil
    \includegraphics[width=0.19\linewidth]{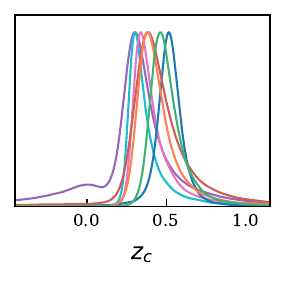} \hfil
    \includegraphics[width=0.19\linewidth]{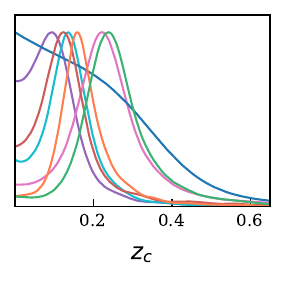} \hfil
    \includegraphics[width=0.19\linewidth]{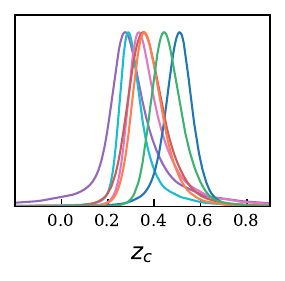} \hfil
    \raisebox{11mm}{\includegraphics[width=0.19\linewidth]{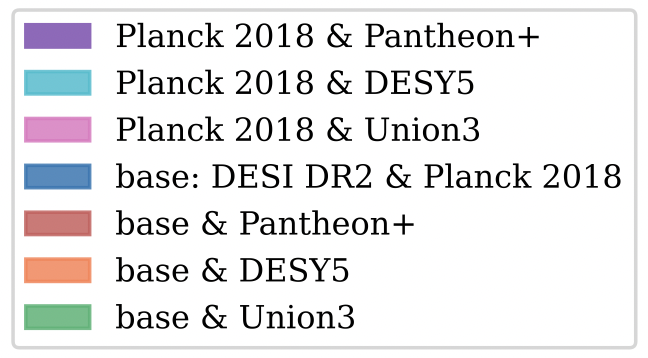} \hfil}
    \\
    [-2pt]
    \parbox[t]{0.19\linewidth}{\centering (a) CPL $\rightarrow -\Lambda$} \hfil
    \parbox[t]{0.19\linewidth}{\centering (b) sCPL}\hfil
    \parbox[t]{0.19\linewidth}{\centering (c) CPL$_{>a_{\rm c}}$}\hfil
    \parbox[t]{0.19\linewidth}{\centering (d) CPL}\hfil
    \parbox[t]{0.19\linewidth}{\centering \,\,\,\,\,\,\,} \hfil\\
    \caption{One-dimensional posterior distributions of the derived parameter $z_{\rm c}$ for the models considered. Colors correspond to the various dataset combinations labeled in~\cref{fig:posteriors_permodel}.}
    \label{fig:zc_permodel}
\end{figure*}

\cref{tab:constraints_main} summarizes the marginalized parameter constraints, reported as posterior means with 68\% credible intervals (CI), for the four models considered: CPL$\rightarrow -\Lambda$ (introduced in~\cref{subsec:modelA}), sCPL (introduced in~\cref{subsec:modelB}), the restricted CPL$_{>a_{\rm c}}$ control case, and the reference CPL model. We also report the derived NECB-crossing redshift $z_{\rm c}=a_{\rm c}^{-1}-1$, where $a_{\rm c}=1+(1+w_0)/w_a$. Near $w_a\simeq 0$, this mapping becomes ill-conditioned and $z_{\rm c}$ may develop strongly non-Gaussian tails; therefore, percentile-based summaries are generally more robust than mean-based estimates. We further quote the best-fit $\Delta$AIC (\cref{eqn:aic}) and Bayes factors (\cref{eqn:bayesfac}) to assess model performance relative to the reference CPL.

Triangular posterior plots for the free parameters are shown in~\cref{fig:posteriors_permodel,fig:posteriors_perdata}, illustrating the impact of each dataset combination on individual models and the relative performance of different models under the same dataset combination, respectively. The corresponding one-dimensional posteriors of the derived parameter $z_{\rm c}$ are displayed in~\cref{fig:zc_permodel,fig:zc_perdata}. The following subsections highlight the main trends in the parameter constraints for each model and discuss the principal findings of our analysis.

\subsection{Baseline CPL}
\label{subsec:baseline_cpl}

We first examine the impact of different dataset combinations on the inferred cosmological parameters. When DESI BAO data are included, our results are broadly consistent with those reported in Tab.~V of the DESI DR2 analysis~\cite{DESI:2025zgx}, with minor differences arising from our use of Planck 2018 measurements alone for CMB anisotropies and the updated Dovekie likelihood~\cite{Popovic:2025glk} for DESY5. In addition, we present results obtained from CMB and SNeIa combinations without BAO data, which allow us to isolate parameter tendencies driven specifically by BAO information.

Planck-only constraints on $(w_0,\,w_a)$ remain broad and largely degeneracy dominated, permitting a wide range of $(H_0,\,\Omega_{\rm m0})$. In the base combination, where DESI BAO data are added to Planck CMB, the CPL model favors a more dynamical region with $w_0=-0.44\pm0.21$ and $w_a=-1.66\pm0.59$, accompanied by lower $H_0$ and higher $\Omega_{\rm m0}$ compared to SNeIa-only combinations. Including low-redshift SNeIa distance data (Pantheon+, DESY5, Union3) significantly tightens the CPL constraints and correspondingly reduces parameter uncertainties in the extended models. With or without DESI BAO, analyses incorporating SNeIa data generally yield higher $H_0$ and lower $\Omega_{\rm m0}$ relative to the base (Planck + DESI) combination. These trends are illustrated in~\cref{fig:posteriors_permodel}\textcolor{blue}{(d)}.

Overall, we observe a relatively well-defined preferred range for the NECB-crossing redshift $z_{\rm c}$ (\cref{fig:zc_permodel}\textcolor{blue}{(d)}). Aside from the Planck-only case, which remains weakly constrained, the CPL model tends to predict crossing within the redshift interval probed by current observations. If CPL provides the best effective description of the data, this suggests that the inferred crossing captures a feature of the reconstructed DE density evolution from~\cref{eq:cpl_rho} in that redshift range. Whether this behavior reflects a genuine physical transition or instead arises as an artifact of the CPL parametrization remains an open question, motivating the alternative constructions explored in this work.

\subsection{CPL$_{>a_{\rm c}}$ (restricted-CPL control case)}
\label{subsec:cpl_ac_results}

The CPL$_{>a_{\rm c}}$ case restricts the CPL parameter space by replacing the dynamical CPL evolution with a cosmological constant at the location of a NECB-crossing, provided that $a_{\rm c} \in [0,1]$. This restriction is strongly disfavored for DESI-inclusive datasets, as reflected by large penalties of $\Delta{\rm AIC}\simeq 11$ for the base combination and $\Delta{\rm AIC}\simeq 5$--$9$ for the base\,+\,SNeIa combinations. 
When DESI BAO data are excluded, however, the fits improve substantially, even though the corresponding parameter constraints do not change dramatically. This behavior indicates that the preference for NECB-crossing is primarily driven by the DESI BAO information. In the absence of DESI, the data no longer favor the region $w(a)<-1$, and the tension with the CPL$_{>a_{\rm c}}$ construction---which forbids NECB-crossing by design---is effectively removed. As a result, the restricted model becomes compatible with the remaining datasets and yields improved goodness of fit.

Another noteworthy feature emerges in the constraints on $w_a$. Across all dataset combinations, an upper bound of $w_a \lesssim -1.60$ at 68\% confidence level is required, enforcing a strictly attenuating DE density toward the past. Since NECB-crossing is forbidden by construction in this model, the slowing-down effect preferred by the expansion history is instead accommodated by reaching the NECB earlier than in the CPL case, as illustrated in~\cref{fig:zc_perdata}. This mechanism, however, comes at a cost. Such early phantom-limiting behavior is not well supported by the data, as reflected by the systematically worse $\Delta{\rm AIC}$ values.
This tension arises because the DE density is forced to remain constant for $a<a_{\rm c}$, yielding an energy density that is lower than its present-day value but still insufficiently suppressed at intermediate redshifts $z \sim 0.5$--$1$, where a reduced expansion rate is preferred. The authors of Ref.~\cite{Ozulker:2025ehg} similarly note that the constraints on $w_a$ are largely prior dominated, particularly at the lower boundary, and that extending the prior range does not lead to improved fits. In addition, we find that allowing more negative values of $w_a$ would merely push $a_{\rm c}$ further outside the physical interval $[0,1]$, causing those regions of parameter space to be effectively excluded from the analysis.

Overall, the CPL$_{>a_{\rm c}}$ construction provides an essential baseline for interpreting our subsequent results for the CPL$\rightarrow -\Lambda$ model, in which the energy density of the cosmological constant phase is taken to be negative. This progression allows us to disentangle the effects of introducing a negative DE density from those arising solely due to the imposed restriction on the CPL evolution.

\subsection{CPL $\rightarrow -\Lambda$ (NECB-triggered sign-switch)}
\label{subsec:modelA_results}

\begin{figure}[t!]
    \includegraphics[width=1.0\columnwidth]{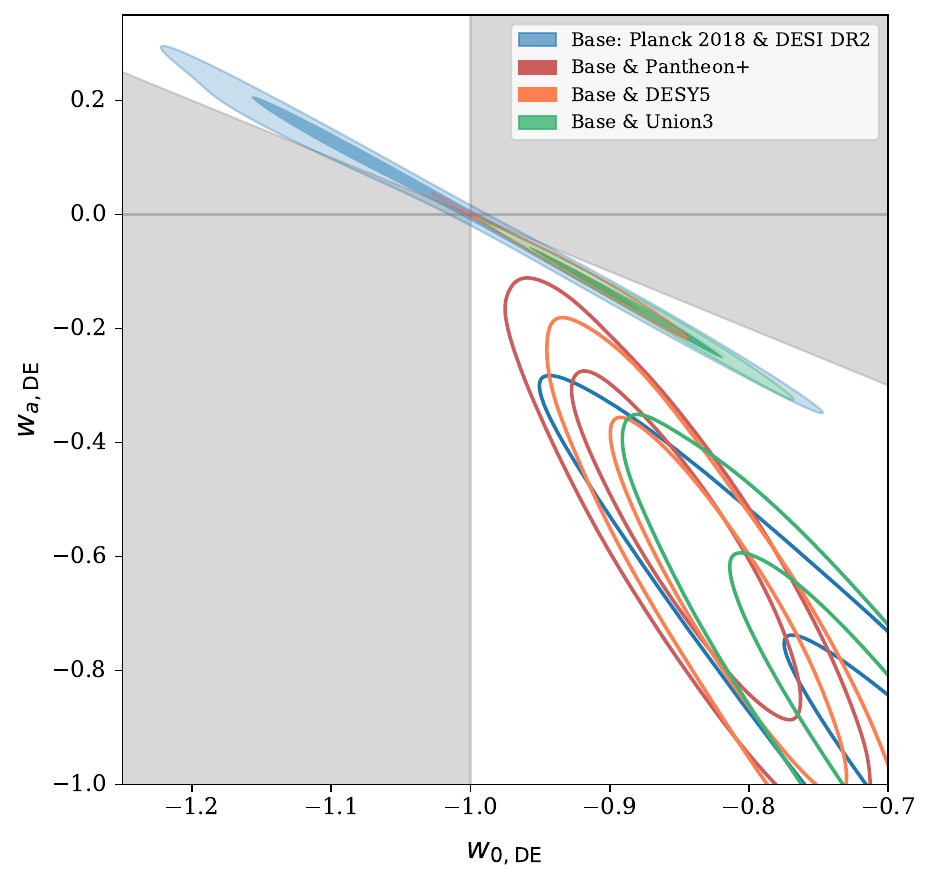}
    \caption{Comparison of dark energy parameter constraints for CPL$\rightarrow -\Lambda$ (filled contours) and the reference CPL model (solid contours). Shaded regions indicate scenarios with $w_{\rm DE}<-1$ or $w_{\rm DE}>-1$ at all times, corresponding to the absence of a transition within $a\in[0,1]$. In these regions, the model effectively reduces to CPL.}
    \label{fig:negcross_w0wa}
\end{figure}

At first glance, the CPL$\rightarrow -\Lambda$ model (introduced in~\cref{subsec:modelA}) appears to yield only modest improvements. However, a closer inspection of the DE parameters reveals a more informative picture. In all dataset combinations, the $(w_0,w_a)$ constraints are driven systematically toward their $\Lambda$ limits, namely $w_0\rightarrow -1$ and $w_a\rightarrow 0$. For instance, for the base (Planck\,+\,DESI) combination we obtain $w_0=-0.99\pm0.11$ and $w_a=-0.02\pm0.15$, in stark contrast to the reference CPL constraints $w_0=-0.44\pm0.21$ and $w_a=-1.66\pm0.59$.
This behavior is illustrated in~\cref{fig:negcross_w0wa}, where we compare the DE EoS parameter constraints of CPL and CPL$\rightarrow -\Lambda$. The introduction of a negative DE density phase in the past effectively tilts the reconstructed $w(z)$ evolution, forcing the model to mimic a less dynamical, nearly constant equation of state parameter in order to remain compatible with the data. In this sense, the apparent shift toward $\Lambda$-like values is not a genuine preference of the data, but rather a consequence of the imposed negative-density phase compensating for the expansion history otherwise achieved through phantom-like dynamics in CPL.
Despite this enforced convergence toward $\Lambda$-like behavior, the preference for a genuinely dynamical EoS remains evident in the Bayesian model comparison. For analyses including DESI BAO, CPL is strongly favored over CPL$\rightarrow -\Lambda$, while for Planck\,+\,SNeIa-only combinations the evidence ranges from mild to strong in favor of CPL.

Moreover, the constraints on both $w_0$ and $w_a$, and in particular on $w_a$, become significantly tighter (\cref{fig:posteriors_perdata}), while the contours for $H_0$ and $\Omega_{\rm m0}$ remain comparably broad. We first note that the forbidden regions with $a_{\rm c}\notin[0,1]$, shaded in~\cref{fig:negcross_w0wa}, play only a minor role in this tightening, as the 95\% confidence contours barely approach these boundaries. Instead, fixing the DE density to a negative $\Lambda$ phase at intermediate-to-high redshifts imposes a more constrained expansion history, directly restricting the range of $(w_0,w_a)$ combinations that can reproduce the observed distance measures. This effect is therefore the primary driver of the tighter posteriors.
The tightening is further enhanced by the data preference for a specific range of the derived parameter $a_{\rm c}$, which sets the location of the DE density sign switch in the CPL$\rightarrow -\Lambda$ model. As shown in~\cref{fig:zc_permodel}\textcolor{blue}{(a)}, the inferred values of $z_{\rm c}$ remain consistent across the two main classes of dataset combinations, namely CMB\,+\,SNeIa and CMB\,+\,BAO\,+\,SNeIa. This behavior contrasts with that of other models, either lacking a sign switch altogether or featuring an additional free parameter controlling it.
Notably, we find $z_{\rm c} \gtrsim 2.46$ for all combinations that include DESI BAO. This is not accidental. The CPL$\rightarrow -\Lambda$ model exhibits a sharp reduction in $\chi^2_{\rm DESI}$ immediately beyond $z=2.33$, corresponding to the highest effective redshift probed by DESI BAO. In other words, negative DE densities are strongly disfavored within the redshift range directly constrained by BAO measurements.
Finally, \cref{fig:zc_perdata} shows that the $z_{\rm c}$ posteriors are relatively broad for this model, which is reflected in the tight constraints on $w_a$. Although $w_a$ is the primary free parameter and $z_{\rm c}$ is derived, the combined evidence indicates that the tightening of the DE EoS constraints ultimately originates from the data preference for a restricted range of $z_{\rm c}$.

\begin{table}[t!]
\caption{Significance of deviations from $w_0=-1$ and $w_a=0$, expressed in terms of equivalent Gaussian standard deviations $\sigma$. The one-dimensional significances are reported as $N_{w_0}$ and $N_{w_a}$, while the joint two-dimensional significance is denoted by $N_\Lambda$. For a detailed discussion of one- and two-dimensional significances, see the Appendix (\cref{sec:appendix}).}
\label{tab:tension_LCDM}
\centering
\begin{tblr}{
  colspec = {l l c c c},
  row{1}  = {font=\bfseries},
}
\toprule
Model & Dataset & $N_{w_0}$ & $N_{w_a}$ & $N_{\Lambda}$ \\
\midrule
CPL$\rightarrow -\Lambda$ & Planck & $1.9\,\sigma$ & $1.9\,\sigma$ & $2.7\,\sigma$ \\
CPL$\rightarrow -\Lambda$ & Planck+Pantheon+ & $2.1\,\sigma$ & $2.1\,\sigma$ & $2.3\,\sigma$ \\
CPL$\rightarrow -\Lambda$ & Planck+DESY5 & $2.5\,\sigma$ & $2.5\,\sigma$ & $2.5\,\sigma$ \\
CPL$\rightarrow -\Lambda$ & Planck+Union3 & $2.9\,\sigma$ & $2.9\,\sigma$ & $2.9\,\sigma$ \\
CPL$\rightarrow -\Lambda$ & Base (Planck+DESI) & $0.1\,\sigma$ & $0.1\,\sigma$ & $0.8\,\sigma$ \\
CPL$\rightarrow -\Lambda$ & Base+Pantheon+     & $2.3\,\sigma$ & $2.3\,\sigma$ & $2.3\,\sigma$ \\
CPL$\rightarrow -\Lambda$ & Base+DESY5         & $2.6\,\sigma$ & $2.6\,\sigma$ & $2.6\,\sigma$ \\
CPL$\rightarrow -\Lambda$ & Base+Union3        & $2.7\,\sigma$ & $2.7\,\sigma$ & $2.6\,\sigma$ \\
\midrule
sCPL & Planck & $0.4\,\sigma$ & $0.7\,\sigma$ & $2.3\,\sigma$ \\
sCPL & Planck+Pantheon+ & $1.6\,\sigma$ & $0.9\,\sigma$ & $2.0\,\sigma$ \\
sCPL & Planck+DESY5 & $2.4\,\sigma$ & $1.7\,\sigma$ & $2.7\,\sigma$ \\
sCPL & Planck+Union3 & $2.7\,\sigma$ & $2.0\,\sigma$ & $3.0\,\sigma$ \\
sCPL & Base (Planck+DESI) & $2.1\,\sigma$ & $2.2\,\sigma$ & $2.2\,\sigma$ \\
sCPL & Base+Pantheon+     & $2.2\,\sigma$ & $1.8\,\sigma$ & $2.3\,\sigma$ \\
sCPL & Base+DESY5         & $2.8\,\sigma$ & $2.3\,\sigma$ & $2.8\,\sigma$ \\
sCPL & Base+Union3        & $3.4\,\sigma$ & $3.0\,\sigma$ & $3.3\,\sigma$ \\
\midrule
CPL$_{>a_{\rm c}}$ & Planck & $0.0\,\sigma$ & $0.0\,\sigma$ & $2.5\,\sigma$ \\
CPL$_{>a_{\rm c}}$ & Planck+Pantheon+ & $1.6\,\sigma$ & $1.6\,\sigma$ & $2.1\,\sigma$ \\
CPL$_{>a_{\rm c}}$ & Planck+DESY5 & $2.1\,\sigma$ & $2.1\,\sigma$ & $2.4\,\sigma$ \\
CPL$_{>a_{\rm c}}$ & Planck+Union3 & $2.2\,\sigma$ & $2.2\,\sigma$ & $2.6\,\sigma$ \\
CPL$_{>a_{\rm c}}$ & Base (Planck+DESI) & $1.1\,\sigma$ & $1.1\,\sigma$ & $2.0\,\sigma$ \\
CPL$_{>a_{\rm c}}$ & Base+Pantheon+     & $1.9\,\sigma$ & $1.9\,\sigma$ & $2.4\,\sigma$ \\
CPL$_{>a_{\rm c}}$ & Base+DESY5         & $2.5\,\sigma$ & $2.5\,\sigma$ & $2.9\,\sigma$ \\
CPL$_{>a_{\rm c}}$ & Base+Union3        & $2.6\,\sigma$ & $2.6\,\sigma$ & $3.0\,\sigma$ \\
\midrule
CPL & Planck & $0.3\,\sigma$ & $0.7\,\sigma$ & $2.2\,\sigma$ \\
CPL & Planck+Pantheon+ & $1.3\,\sigma$ & $1.1\,\sigma$ & $1.3\,\sigma$ \\
CPL & Planck+DESY5 & $2.0\,\sigma$ & $1.9\,\sigma$ & $2.0\,\sigma$ \\
CPL & Planck+Union3 & $2.4\,\sigma$ & $2.0\,\sigma$ & $2.5\,\sigma$ \\
CPL & Base (Planck+DESI) & $2.7\,\sigma$ & $3.1\,\sigma$ & $3.3\,\sigma$ \\
CPL & Base+Pantheon+     & $3.0\,\sigma$ & $3.1\,\sigma$ & $3.2\,\sigma$ \\
CPL & Base+DESY5         & $3.5\,\sigma$ & $3.7\,\sigma$ & $3.7\,\sigma$ \\
CPL & Base+Union3        & $3.9\,\sigma$ & $3.8\,\sigma$ & $3.9\,\sigma$ \\
\bottomrule
\end{tblr}
\end{table}

In principle, as $w_0 \rightarrow -1$ and $w_a \rightarrow 0$, tensions with a cosmological constant would be expected to diminish. However, the simultaneous reduction in parameter uncertainties limits the extent to which the inferred significances decrease. To quantify the proximity of our constraints to $\Lambda$, we adopt the null hypothesis $w_0=-1$ and $w_a=0$ and compute the significance of deviations from these values in units of equivalent Gaussian standard deviations $\sigma$. 
We evaluate the one-dimensional significances $N_{w_0}$ and $N_{w_a}$ using a two-tailed test~\cite{Cowan:2010js}, which remains applicable in the presence of non-Gaussian posteriors. For the joint two-dimensional significance $N_\Lambda$, we perform an analogous test in the $(w_0,w_a)$ posterior space and apply Wilks' theorem~\cite{Wilks:1938dza} to express the probability of obtaining a non-null result in terms of Gaussian-equivalent sigmas. The resulting values are reported in~\cref{tab:tension_LCDM}.
For the base combination, we find $N_\Lambda = 0.8\,\sigma$, indicating that the tension with $\Lambda$ is fully resolved. For the remaining dataset combinations, despite the constraints moving closer to the cosmological constant limit and their associated significances being reduced, $(w_0,w_a)$ remain mildly displaced from $\Lambda$, with $N_\Lambda$ in the range $2.3\,\sigma$--$2.9\,\sigma$.

Unfortunately, the CPL$\rightarrow -\Lambda$ model does not improve the overall fit to the data. The corresponding $\Delta$AIC values are positive for all dataset combinations, indicating worse performance than CPL, since both models have the same number of free parameters. This degradation is more pronounced for DESI-inclusive analyses. This behavior can be understood as a consequence of the model construction: both CPL$\rightarrow -\Lambda$ and CPL$_{>a_{\rm c}}$ exclude a pure cosmological constant as a viable realization within their CPL-parametrized regime. This interpretation is supported by the systematic shifts observed in the $(w_0,w_a)$ constraints. In particular, as $w_a\rightarrow 0$, the crossing scale factor $a_{\rm c}\rightarrow\infty$, and the transition to the (negative) cosmological constant phase never occurs within the physically relevant redshift range. Enforcing $w_a\neq 0$ therefore impacts the inferred constraints, leading to visibly non-Gaussian posteriors in cases where the parameter space approaches $w_a\simeq 0$. This non-Gaussianity is most pronounced for the base combination, where we observe a clear bimodality that warrants a dedicated discussion in~\cref{sec:bimodality}.

Concerning the $H_0$ tension, the CPL$\rightarrow -\Lambda$ model does not substantially outperform the other scenarios. Nevertheless, an interesting exception arises for the base (Planck\,+\,DESI) dataset combination. While CPL$_{>a_{\rm c}}$ already raises the CPL value of $H_0=63.8^{+1.7}_{-2.1}\,{\rm km\,s^{-1}\,Mpc^{-1}}$ to $H_0=67.0^{+1.7}_{-1.5}\,{\rm km\,s^{-1}\,Mpc^{-1}}$, the inclusion of a negative DE density phase in CPL$\rightarrow -\Lambda$ further increases the inferred value to $H_0=68.5\pm1.8\,{\rm km\,s^{-1}\,Mpc^{-1}}$, the highest among the models considered. Although this shift remains insufficient to reconcile the inferred value with the local distance ladder measurement, $H_0^{\rm LD}=73.50\pm0.81\,{\rm km\,s^{-1}\,Mpc^{-1}}$~\cite{H0DN:2025lyy,Riess:2021jrx}, the results obtained with CPL$\rightarrow -\Lambda$ nevertheless provide valuable insight into how modifications of the DE sector influence late-time expansion and point toward directions for future model building.

\subsection{sCPL (free sign-switch in $\rho_{\rm DE}$)}
\label{subsec:modelB_results}

We introduced the sCPL model in~\cref{subsec:modelB} with a free sign-switch redshift $z_\dagger$ in order to superpose the dynamical CPL EoS with a sign-switching feature in a more flexible manner. Unlike CPL$\rightarrow -\Lambda$, the sCPL construction admits a limit in which it reduces to the $\Lambda_{\rm s}$CDM model~\cite{Akarsu:2019hmw,Akarsu:2021fol,Akarsu:2022typ,Akarsu:2023mfb}. 
A key result for sCPL is that, across all dataset combinations, the inferred posteriors closely track those of CPL for DESI-inclusive analyses and differ only mildly when DESI BAO data are excluded. When the lower bound on $z_\dagger$ exceeds the maximum redshift probed by a given dataset, the analysis becomes insensitive to the presence of the sign switch, rendering sCPL effectively indistinguishable from CPL. In particular, DESI BAO data drive $z_\dagger \gtrsim 2.38$, placing the transition entirely beyond the redshift range probed by BAO measurements. In contrast, SNeIa data alone allow the switch to occur, but typically near the upper end of the prior range specified in~\cref{sec:data_and_meth}, reflecting their more limited constraining power.
We note that the upper prior bound on $z_\dagger$ lies relatively close to the inferred lower limits. While standard practice would suggest extending the prior range, doing so would be inconsequential in this case. The posteriors plateau once the transition is pushed beyond the observationally accessible redshift range, where no data are available to break the degeneracy between CPL and sCPL.

Despite yielding parameter constraints that closely resemble those of CPL, the sCPL model incurs larger $\Delta$AIC values due to the penalty associated with its additional free parameter. For the Planck\,+\,SNeIa combinations, sCPL therefore fits the data slightly worse than CPL. These degraded $\Delta$AIC values can be partially attributed to the ``sandwich'' behavior permitted by the model. Across all dataset combinations, the preferred solutions correspond to a $p$-phantom $\to p$-quintessence $\to n$-phantom evolution. We also note that sCPL is subject to the same limitations as CPL, namely the abundance of low-redshift data relative to intermediate-redshift information, as well as reconstruction artifacts inherent to a linear EoS parametrization. In the absence of higher-redshift BAO measurements, it remains difficult to determine whether CPL is favored because it effectively mimics a slowly sign-switching DE~\cite{Akarsu:2025gwi} or because the DE EoS is genuinely dynamical.

Furthermore, while the DE parameters retain their preference for dynamical behavior, the inclusion of negative DE densities at early times leads only to modest increases in the inferred values of $H_0$. As a result, the anticipated alleviation of the $H_0$ tension is not realized. Nevertheless, the sCPL model remains valuable for interpreting the results obtained with CPL$\rightarrow -\Lambda$. In particular, when comparing the constraints on $z_\dagger$ in sCPL with the derived $z_{\rm c}$ constraints in CPL$\rightarrow -\Lambda$, we find a striking similarity between the two. Since $z_\dagger$ and $z_{\rm c}$ respectively mark the redshifts at which the transition to negative DE densities occurs, this correspondence suggests that the location of the sign switch plays a decisive role in shaping the $(w_0,w_a)$ constraints in the CPL$\rightarrow -\Lambda$ model.
This observation also helps explain why the inferred $z_{\rm c}$ values in CPL$\rightarrow -\Lambda$ differ markedly from those of the other models. The DESI BAO data exhibit a strong disfavoring of negative DE densities within their effective redshift range, to the extent that the model selects less favorable $(w_0,w_a)$ values rather than allowing a sign switch to occur where BAO measurements are active. With these considerations in mind, we now proceed to the next section, where we elaborate on one of the core findings of this work.

\subsection{Bimodality in the base constraints of CPL$\rightarrow -\Lambda$}
\label{sec:bimodality}

As noted earlier, in the limit $w_a \rightarrow 0$ the crossing scale factor $a_{\rm c}$ becomes ill defined. For models in which $a_{\rm c}$ plays a central role in characterizing the DE behavior—namely CPL$\rightarrow -\Lambda$ and CPL$_{>a_{\rm c}}$—this leads to non-Gaussian features in the posterior distributions (\cref{fig:posteriors_permodel}\textcolor{blue}{(a)} and \textcolor{blue}{(c)}). In dataset combinations including SNeIa, these deviations are relatively mild and primarily appear as skewness and extended tails. In contrast, for the base (Planck\,+\,DESI) combination, we observe highly irregular posteriors for CPL$_{>a_{\rm c}}$ and a pronounced bimodality with nearly resolved peaks for the CPL$\rightarrow -\Lambda$ model. 
In this section, we investigate the origin of the bimodality observed in the base results for CPL$\rightarrow -\Lambda$ and discuss its cosmological implications.

\begin{figure}[b!t]
    \centering
    \includegraphics[width=1.\columnwidth]{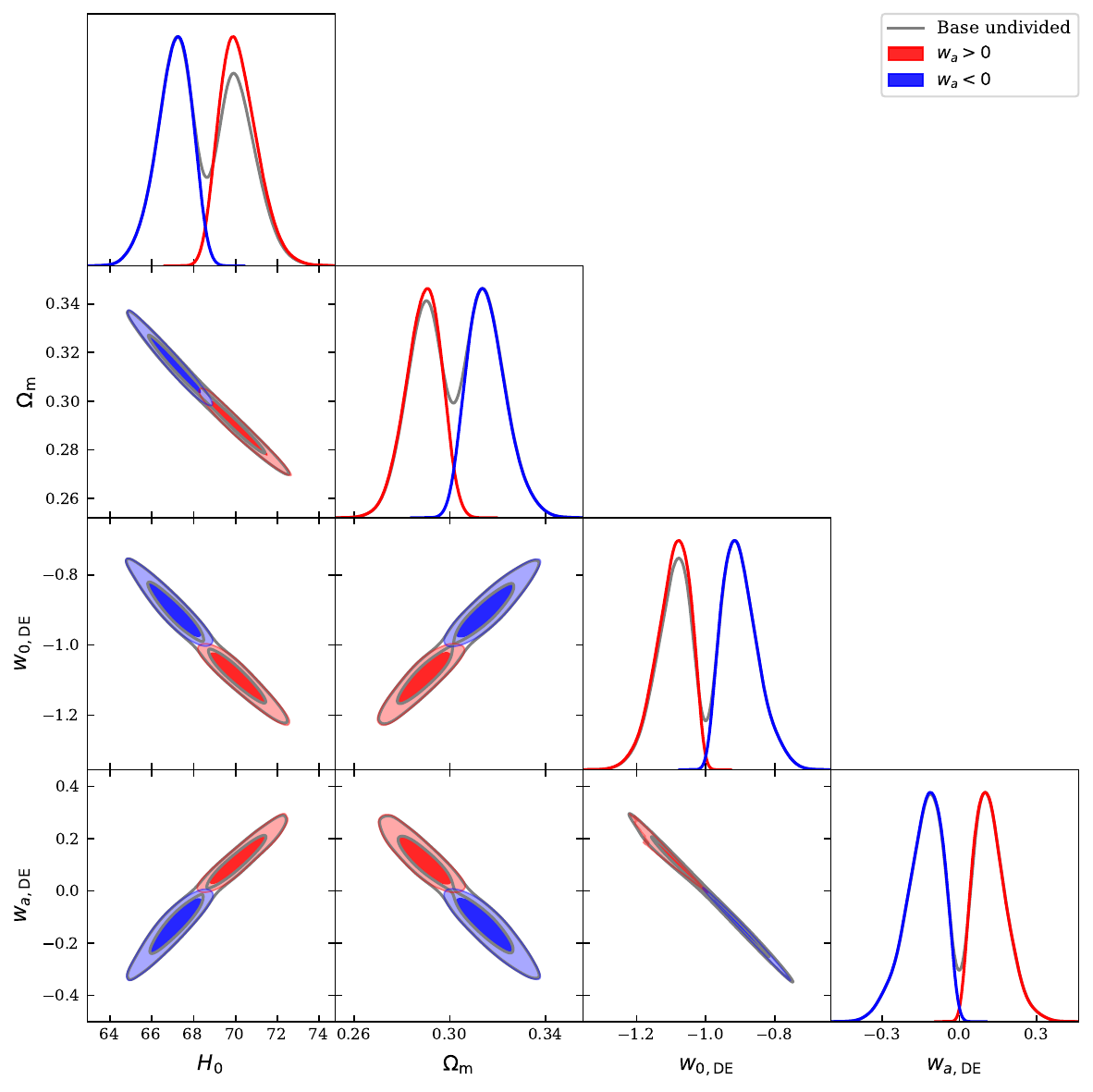}
    \caption{The base (Planck 2018\,+\,DESI DR2) analysis chain is divided at $w_a=0$ into two regions in order to investigate the bimodality observed in the CPL$\rightarrow -\Lambda$ posteriors.}
    \label{fig:bimodal_split_tri}
\end{figure}

\begin{table}[t!]
    \caption{The base (Planck 2018\,+\,DESI DR2) analysis chain is divided at $w_a=0$ into two regions to investigate the bimodality observed in the CPL$\rightarrow -\Lambda$ posteriors. The table reports the posterior mean and the 68\% CI for each region.}
    \centering
    \begin{threeparttable}
    \begin{tabular} { l|  c | c c}
    & \quad  Whole Chain \quad  &  \quad $w_a >0 $ \quad & \, $w_a <0$ \\
    \hline \hline
    {\boldmath$w_0$} &  $-0.99\pm 0.11             $ & $-1.095^{+0.060}_{-0.036}  $ & $-0.899^{+0.038}_{-0.063}  $\\

    {\boldmath$w_a$} & $-0.02\pm 0.15              $ & $0.121^{+0.045}_{-0.077}   $ & $-0.138^{+0.088}_{-0.051}  $\\
    \hline
    $H_0                        $ &  $68.5\pm 1.8                $ & $70.17^{+0.72}_{-1.1}      $ & $67.04^{+0.97}_{-0.69}     $\\
    
    $\Omega_\mathrm{m0}          $ & $0.303\pm 0.016             $ & $0.2888^{+0.0084}_{-0.0065}$ & $0.3157^{+0.0066}_{-0.0094}$\\
    
    $z_{\rm c}$ \textbf{\tnote{*}} & $2.67_{-0.15}^{+1.34}$ & $2.79^{+2.80}_{-0.15}$ & $2.60^{+0.68}_{-0.14}$\\
    
    $\Delta \chi^2_{\rm min}$ \textbf{\tnote{**}} \, \ & $0.0000$ & $0.2766$ & $0.0000$ \\
    \hline\hline
    \end{tabular} %
    \begin{tablenotes}
        \small
        \item[*] For the derived parameter $z_{\rm c}$, we report the posterior peak rather than the mean. For distributions with long tails, the mean may lie outside the 68\% confidence interval, which is the case here.
        \item[**] $\Delta\chi^2_{\rm min}\equiv \chi^2_{\rm min,region}-\chi^2_{\rm min,whole\text{-}chain}$
    \end{tablenotes}
    \end{threeparttable}
    \label{tab:bimodal_split}
\end{table}

\begin{figure}[h!tb]
    \centering
    \includegraphics[width=1.\columnwidth]{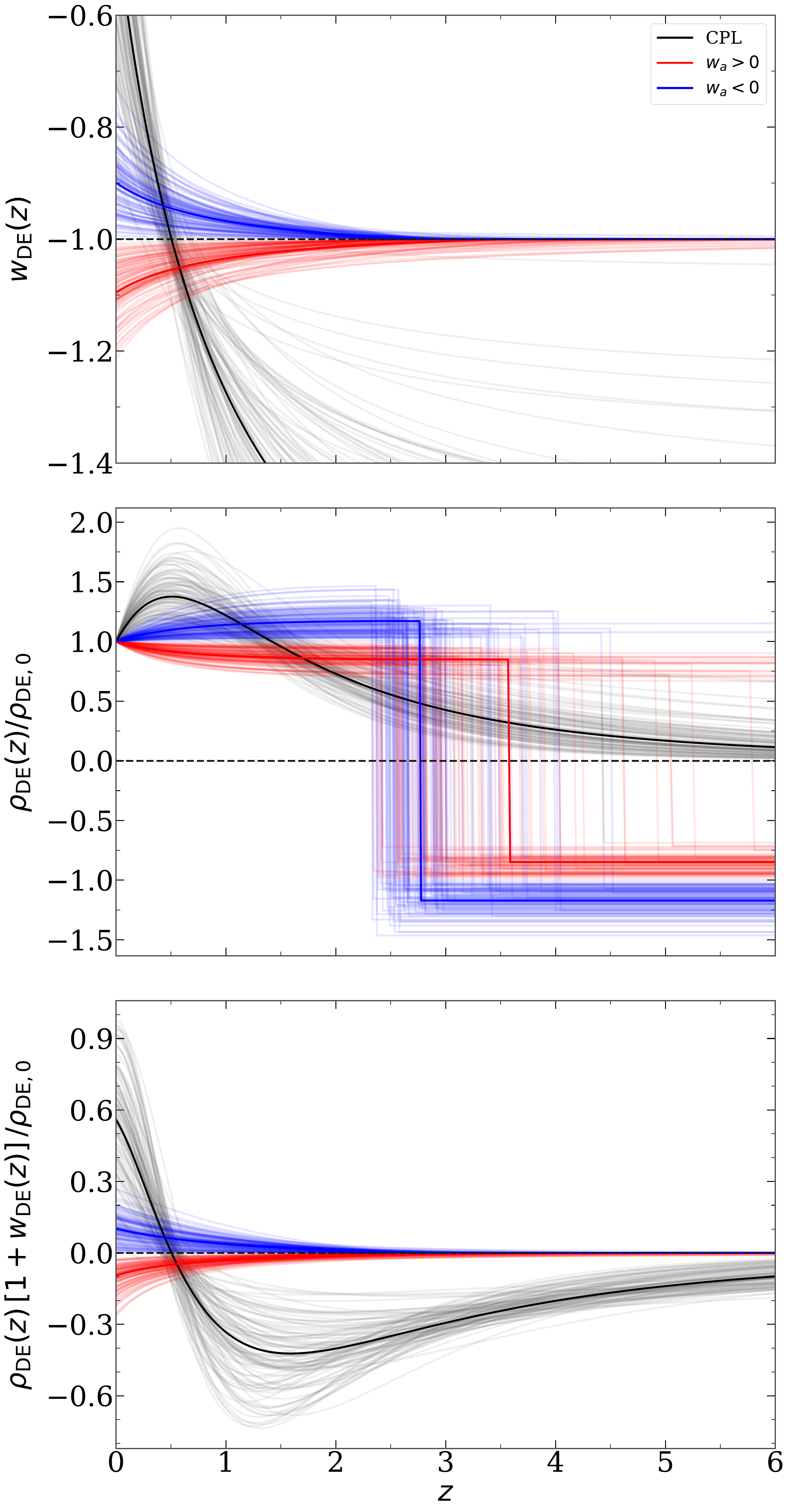}
    \caption{Dark energy EoS evolution (\textbf{top panel}), energy density evolution (\textbf{middle panel}), 
and $\rho_{\rm DE}+p_{\rm DE}$ (\textbf{bottom panel}) for the two regions identified in 
\cref{fig:bimodal_split_tri}, compared with the CPL-parametrized dark energy constraints from the base analysis. 
Multiple evolution histories are generated using the \textbf{fgivenx} package~\cite{Handley_2018}, with fainter 
lines corresponding to samples further from the posterior mean.}
    \label{fig:bimodal_split_evol}
\end{figure}

We begin by separating the two modes of the bimodal posterior in order to study each regime independently. Following standard practice, we impose a cut in post-processing and reweight the chains to implement the split illustrated in~\cref{fig:bimodal_split_tri}. We find that the bimodality observed across all parameters is consistently resolved by a cut at $w_a=0$. The two resulting regions are treated as independent chains, from which we derive the parameter constraints reported in~\cref{tab:bimodal_split}.
The posterior separates into two mirror branches across the $w_a=0$ boundary: a ``red'' mode characterized by $w_0<-1$ and $w_a>0$, and a ``blue'' mode with $w_0>-1$ and $w_a<0$. Despite their different locations in the $(w_0,w_a)$ plane, both branches correspond to comparable NECB-crossing redshifts, $z_{\rm c}\sim3$. From the bottom panel of \cref{fig:bimodal_split_evol}, we identify the red mode as a thawing-like phantom
and the blue mode as a freezing-like quintessence scenario \emph{at late times}, based on which side of the
NECB these evolutions occupy. However, for a sign-switching scenario to be physically realized with a smooth
transition in $\rho_{\rm DE}$, both modes must exhibit phantom behavior at $a=a_{\rm c}$ in order to satisfy the
continuity equation in \cref{eqn:cont_eqn}.
The absence of bimodality in the control CPL$_{>a_{\rm c}}$ case, together with the unimodal base $z_{\rm c}$ posteriors (\cref{fig:posteriors_perdata}\textcolor{blue}{(a)}), indicates that the presence of negative energy densities and the location of the transition to that regime play a central role in shaping the $(w_0,w_a)$ constraints. To illustrate this, we show the corresponding DE EoS and energy density evolutions for the two bimodal branches in~\cref{fig:bimodal_split_evol}. The asymptotic behavior of the EoS in both regimes approaches the $\Lambda$ limit, suggesting that each branch tends toward a $\Lambda_{\rm s}$CDM-like scenario, even though the model cannot reduce to it by construction.
Technically, the model does not forbid $w_a=0$. Rather, it enforces a choice between remaining in a CPL phase with $w_a=0$ or undergoing a transition to a negative-density era. This tension naturally raises the question of why the sCPL model does not converge to its $\Lambda_{\rm s}$CDM limit. While a definitive answer lies beyond the scope of this phenomenological study, we can state with certainty that once a negative cosmological constant phase is imposed in the past, consistency of the CPL branch at $a<a_{\rm c}$ requires the equation of state parameter to approach $w_{\rm DE}=-1$.

\begin{figure}[tb]
    \centering
    \includegraphics[width=1.0\columnwidth]{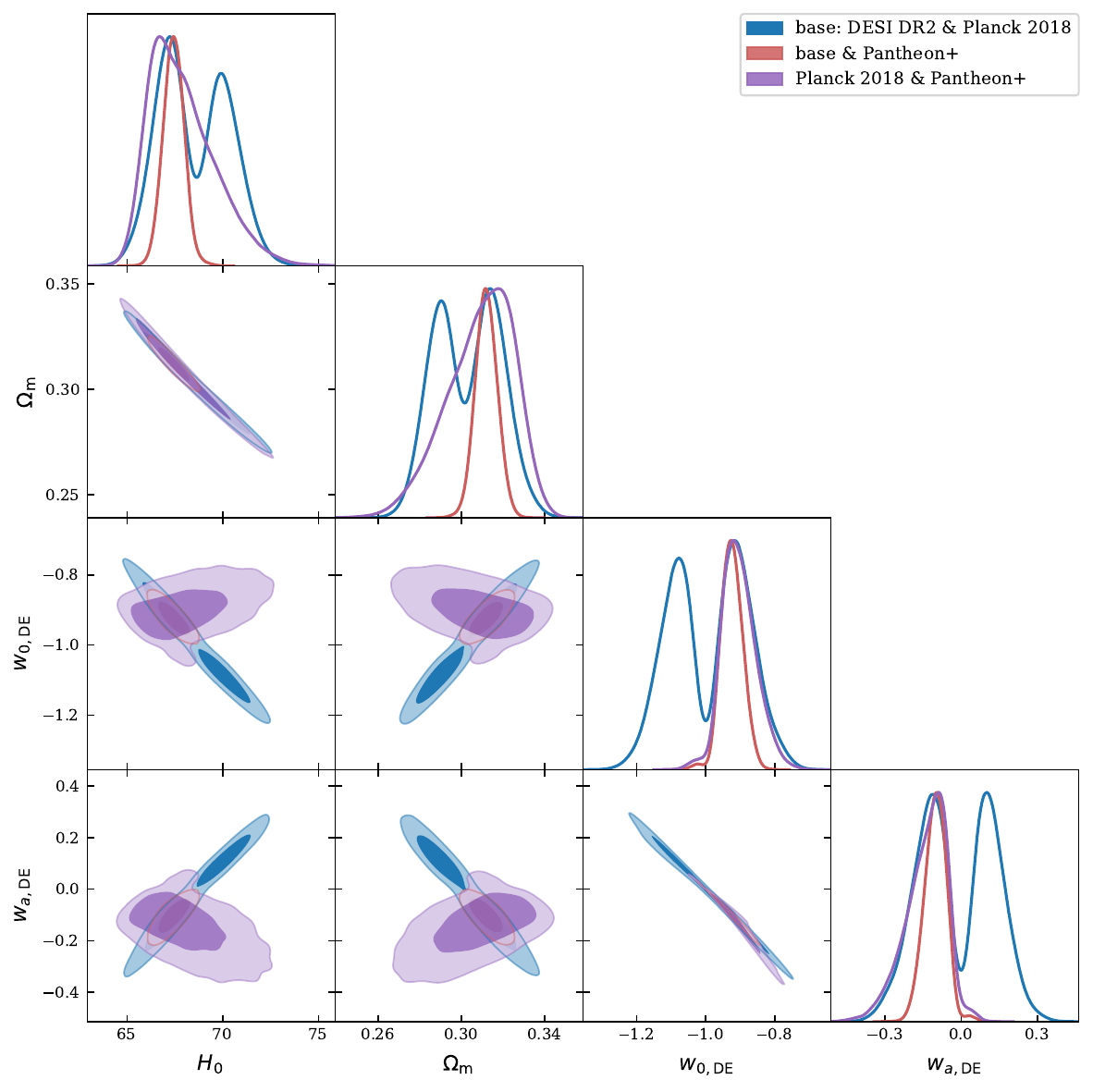}
    \caption{One- and two-dimensional posterior distributions of the CPL$\rightarrow -\Lambda$ model parameters for various combinations of three datasets.}
    \label{fig:bimodality_comp_tri}
\end{figure}

\begin{figure}[t!]
    \centering
    \includegraphics[width=1.0\columnwidth]{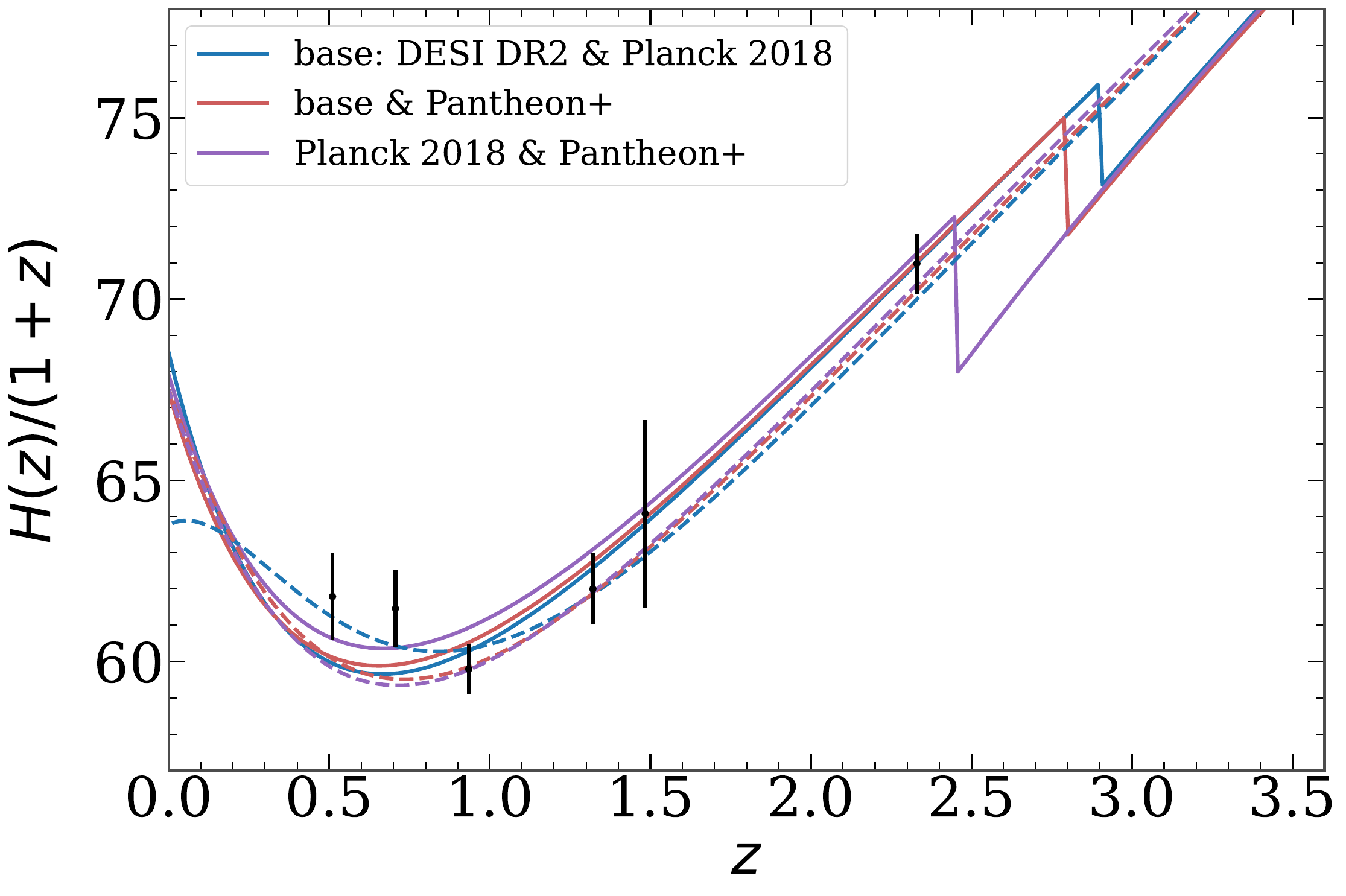}
    \caption{Redshift evolution of the comoving Hubble parameter $\dot{a}=H(z)/(1+z)$ for CPL$\rightarrow -\Lambda$ (solid lines) and CPL (dashed lines). The evolutions are computed using the posterior mean parameter values. Black points correspond to DESI BAO Hubble distance measurements $D_H(z)/r_d$ from Tab.~IV of Ref.~\cite{DESI:2025zgx}, assuming $r_d=147.05\,\mathrm{Mpc}$. Note how the inclusion of SNe data strongly influences which peak of the base posterior is preferred.}
    \label{fig:bimodality_comp_adot}
\end{figure}

\begin{figure*}[h!tb]
    \includegraphics[width=0.33\textwidth]{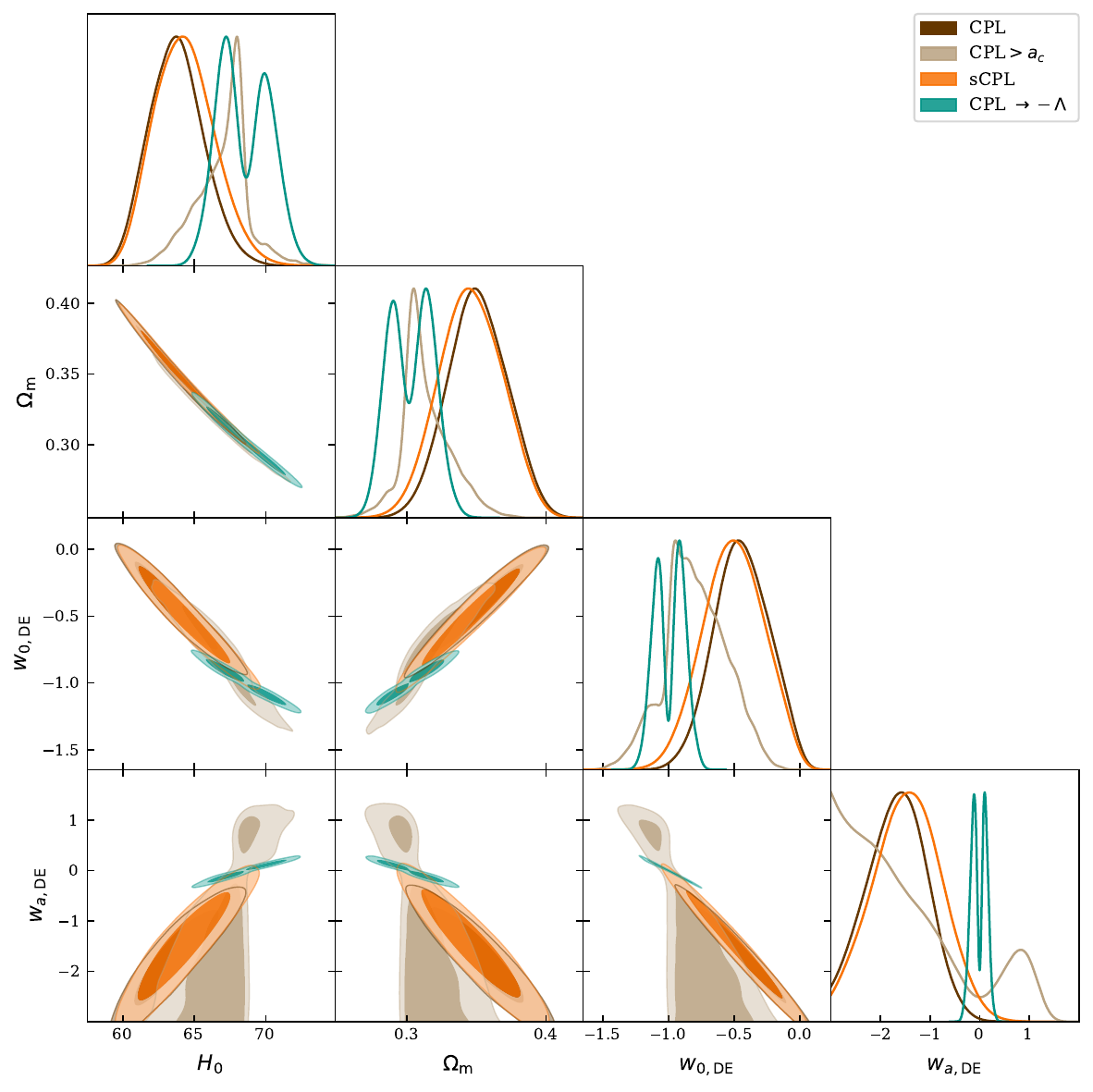}\\
    \parbox[t]{1.0\linewidth}{\centering (a) Base: Planck 2018 \& DESI DR2}\\
    [10pt]
    
    \includegraphics[width=0.33\textwidth]{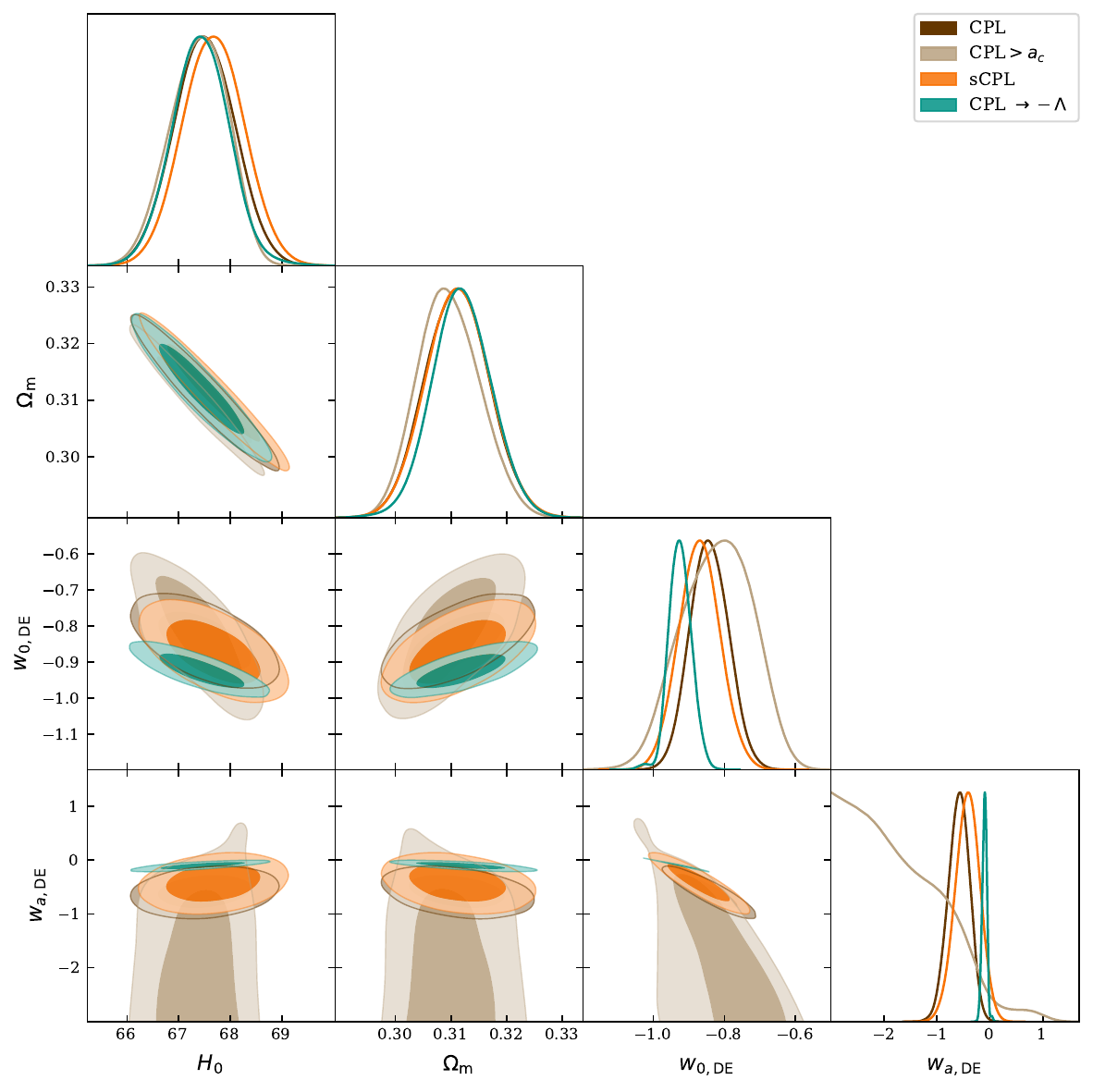}\hfil
    \includegraphics[width=0.33\textwidth]{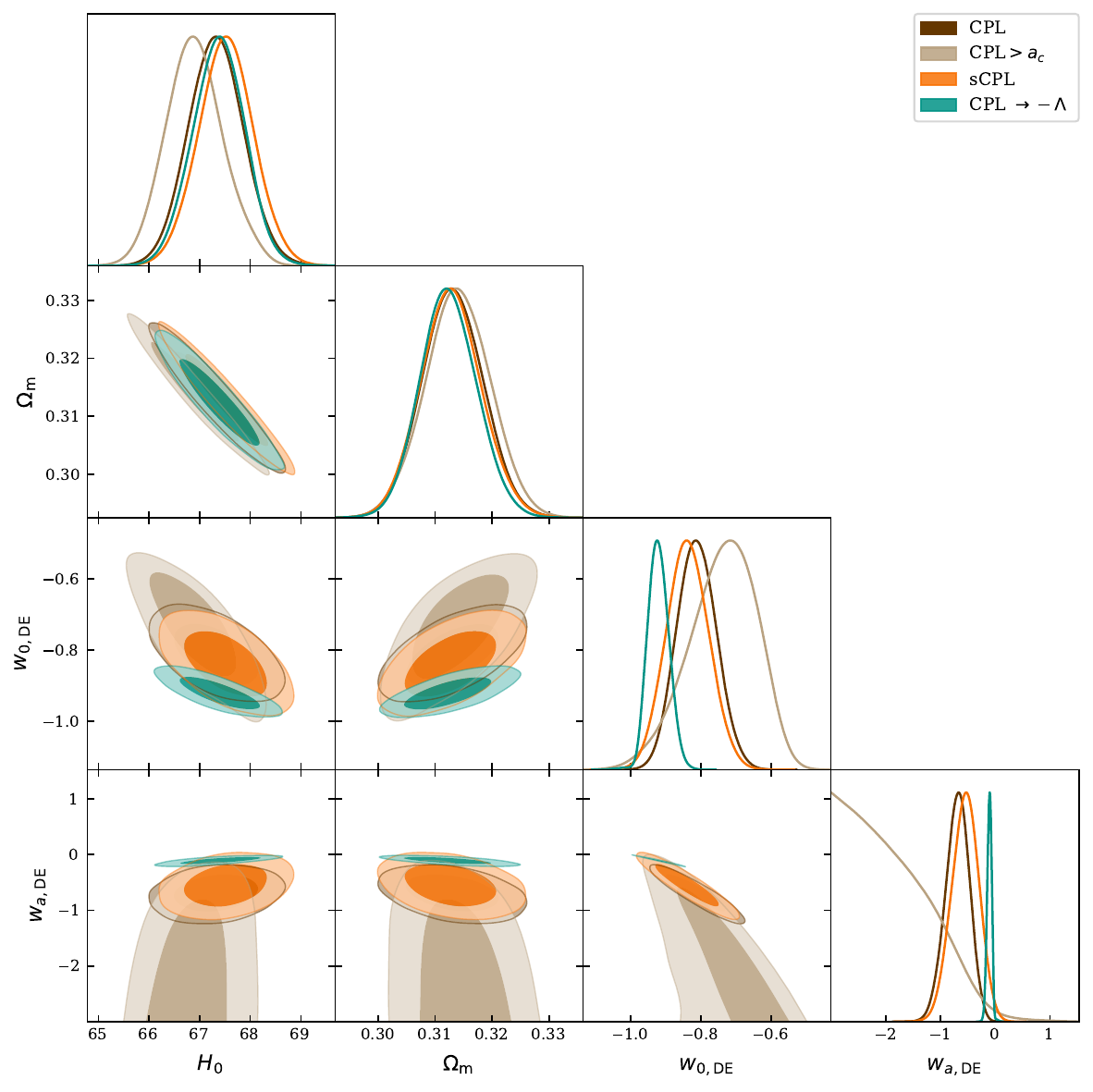}\hfil
    \includegraphics[width=0.33\textwidth]{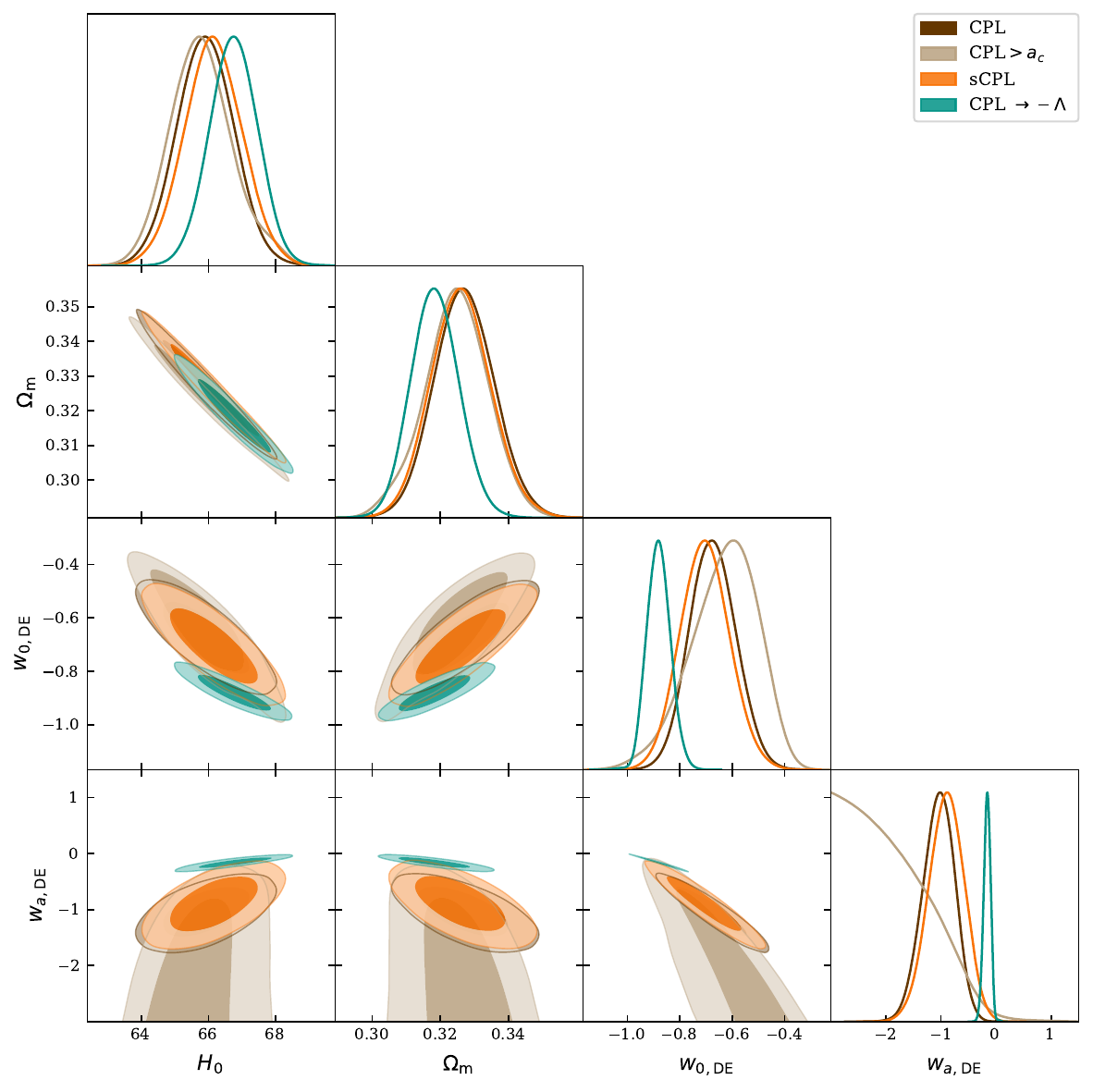}\\
    [-2pt]
    \parbox[t]{0.33\linewidth}{\centering (b) Base \& Pantheon+}\hfill
    \parbox[t]{0.33\linewidth}{\centering (c) Base \& DESY5}\hfill
    \parbox[t]{0.33\linewidth}{\centering (d) Base \& Union3}\\
    [10pt]

    \includegraphics[width=0.33\textwidth]{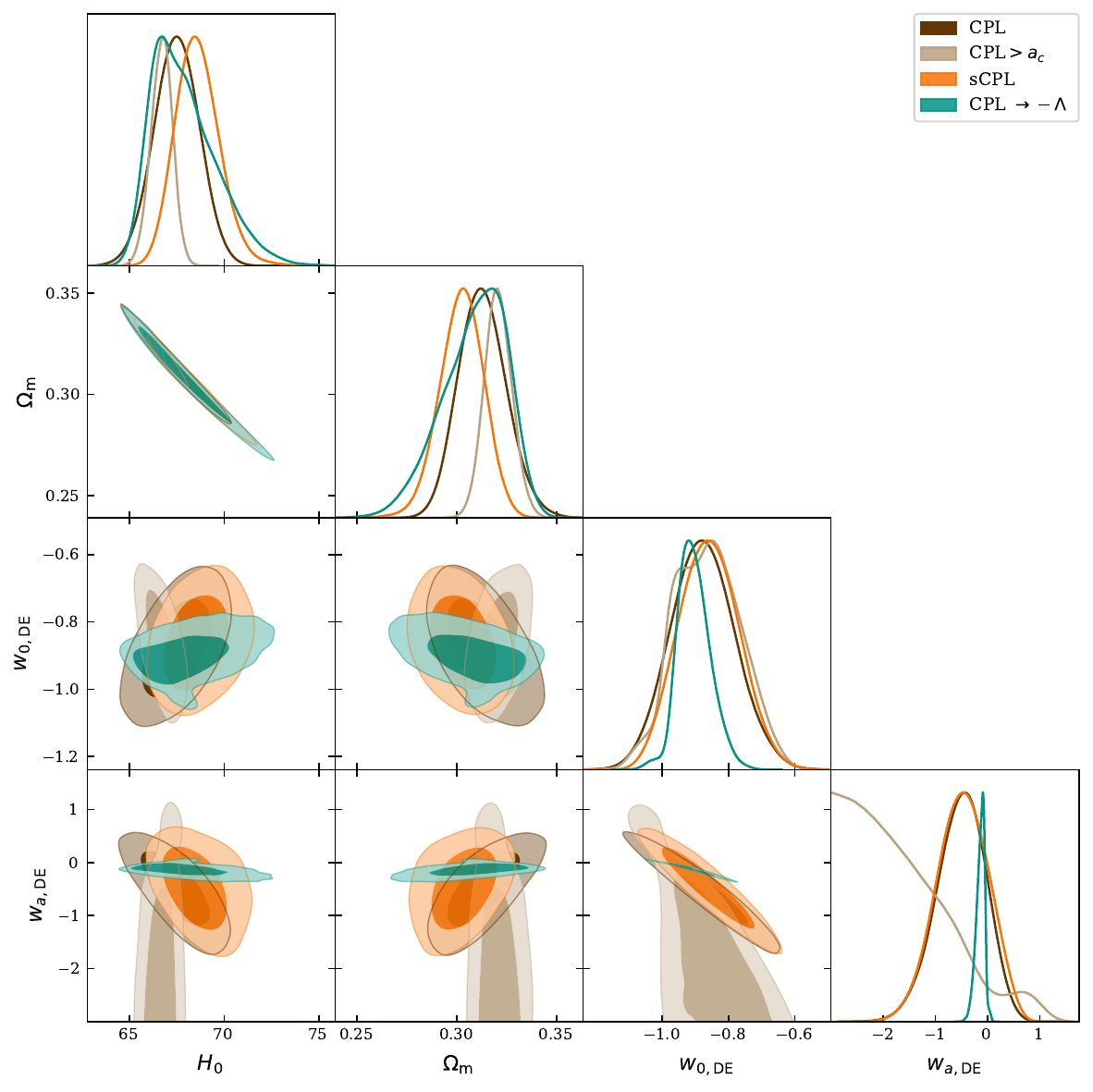}\hfil
    \includegraphics[width=0.33\textwidth]{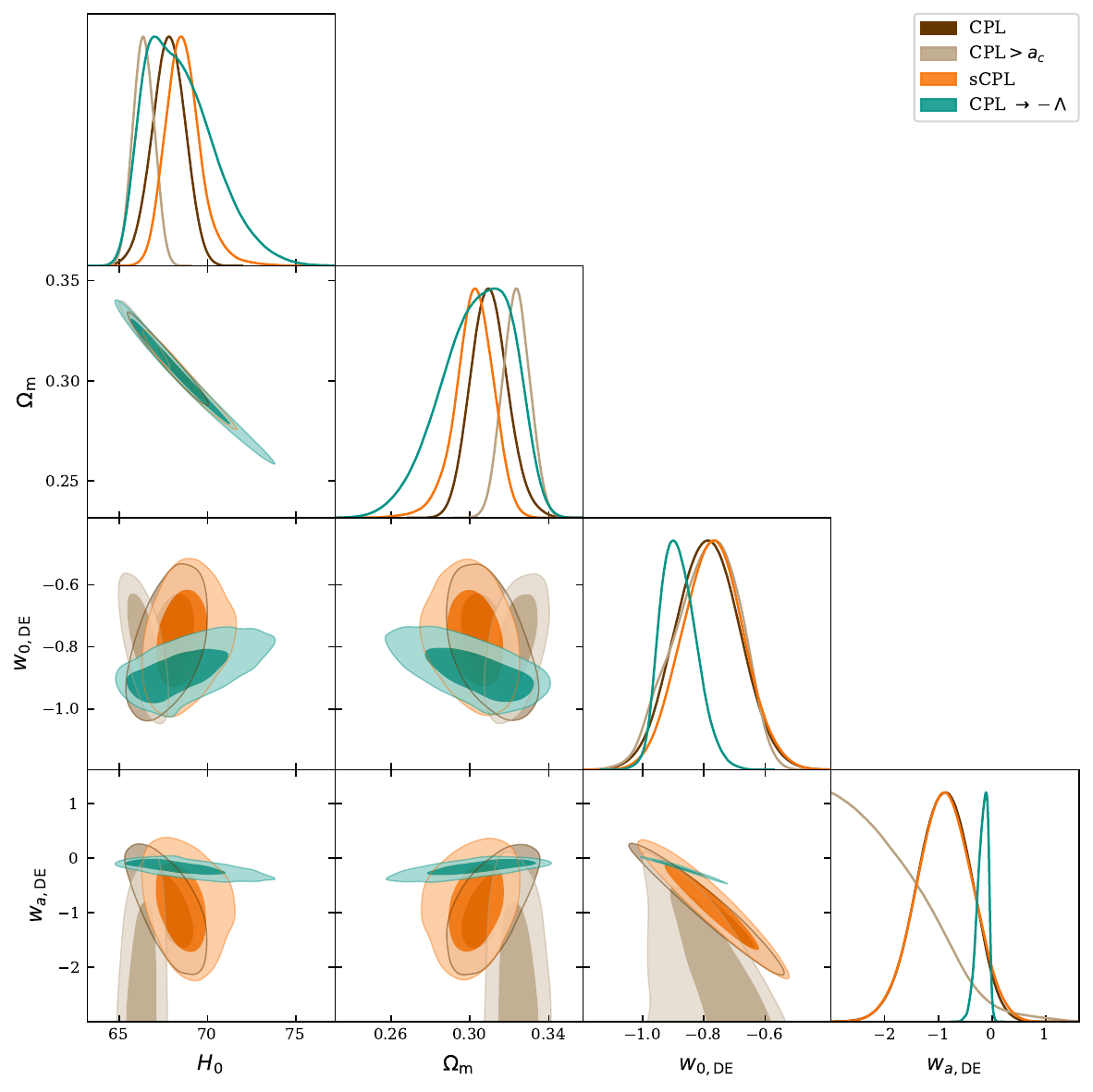}\hfil
    \includegraphics[width=0.33\textwidth]{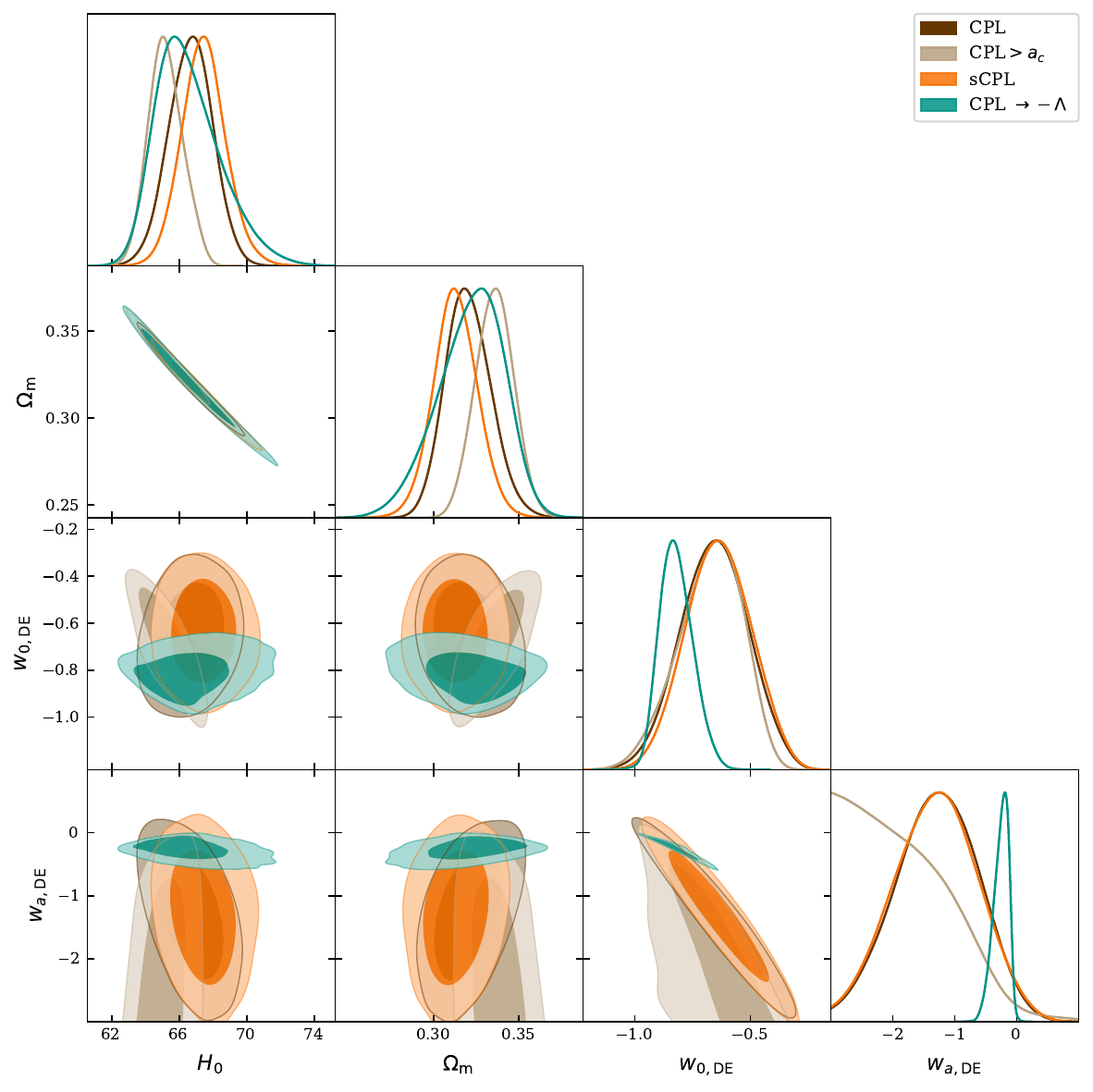}\\
    [-2pt]
    \parbox[t]{0.33\linewidth}{\centering (e) Planck 2018 \& Pantheon+}\hfill
    \parbox[t]{0.33\linewidth}{\centering (f) Planck 2018 \& DESY5}\hfill
    \parbox[t]{0.33\linewidth}{\centering (g) Planck 2018 \& Union3}\\
    [10pt]
     
    \caption{One- and two-dimensional marginalized posterior distributions for the four models, presented for each dataset combination.}
    \label{fig:posteriors_perdata}
\end{figure*}

\begin{figure*}[h!tb]
    \centering
    \includegraphics[width=0.24\textwidth]{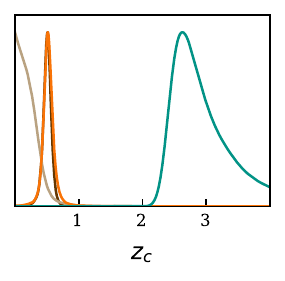}
    \includegraphics[width=0.24\textwidth]{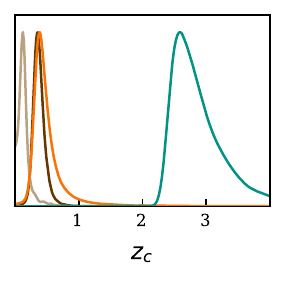}
    \includegraphics[width=0.24\textwidth]{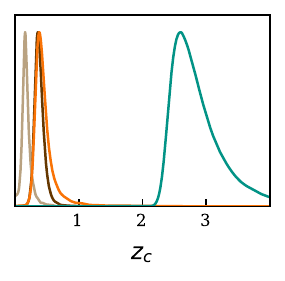}
    \includegraphics[width=0.24\textwidth]{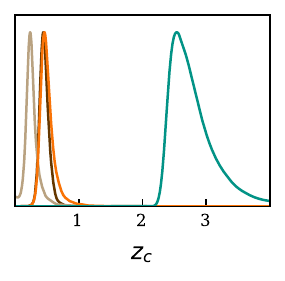}\\
    [-2pt]
    \parbox[t]{0.24\linewidth}{\centering (a) Base} \hfill
    \parbox[t]{0.24\linewidth}{\centering (b) Base \& Pantheon+}\hfill
    \parbox[t]{0.24\linewidth}{\centering (c) Base \& DESY5}\hfill
    \parbox[t]{0.24\linewidth}{\centering (d) Base \& Union3}\\
    [10pt]
    \includegraphics[width=0.24\textwidth]{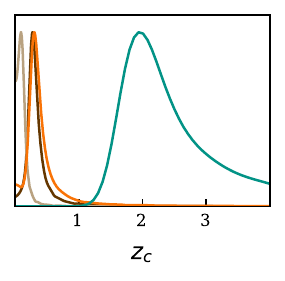}
    \includegraphics[width=0.24\textwidth]{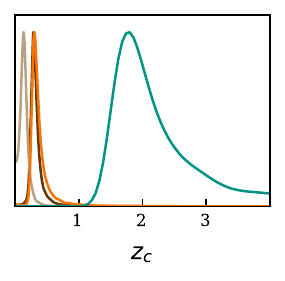}
    \includegraphics[width=0.24\textwidth]{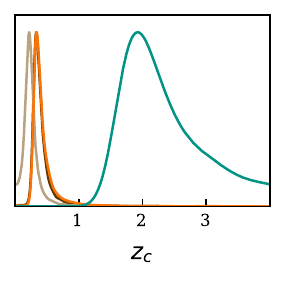}
    \raisebox{11.5mm}{\includegraphics[width=0.24\textwidth]{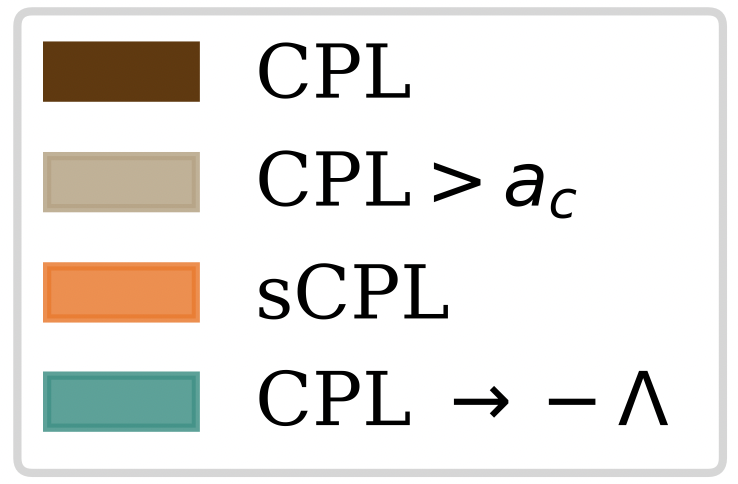}}
    \\
    [-2pt]
    \parbox[t]{0.24\linewidth}{\centering (e) Planck \& Pantheon+}
    \parbox[t]{0.24\linewidth}{\centering (f) Planck \& DESY5}
    \parbox[t]{0.24\linewidth}{\centering (g) Planck \& Union3}
    \parbox[t]{0.24\linewidth}{\centering \,\,\,\,\,}
    \\
    \caption{One-dimensional posterior distributions of the derived parameter $z_{\rm c}$ for the four models, shown separately for each dataset combination. Colors correspond to the models labeled in~\cref{fig:posteriors_perdata}.}
    \label{fig:zc_perdata}
\end{figure*}

We examine the impact of different combinations of Planck, DESI, and Pantheon$+$ data on the posterior distributions in~\cref{fig:bimodality_comp_tri}. The Pantheon$+$ data exhibit a strong preference for the blue ($w_a<0$) branch over the red one, both with and without the inclusion of DESI BAO. As SNeIa provide dense low-redshift distance information, they break the near symmetry between the $(w_0<-1,w_a>0)$ and $(w_0>-1,w_a<0)$ solutions, which can otherwise reproduce integrated high-redshift distance constraints with comparable accuracy.
DESI BAO alone remain largely insensitive to the distinction between the two DE regimes. However, when combined with SNeIa data, the parameter space corresponding to the red ($w_a>0$) branch becomes increasingly suppressed. In contrast to the CPL case, SNeIa-only combinations favor lower values of $H_0$, whereas the base combination allows for higher values, reaching $H_0 = 70.17^{+0.72}_{-1.1}\,\mathrm{km\,s^{-1}\,Mpc^{-1}}$ within the red ($w_a>0$) region.
The resultant expansion rates, viz., comoving Hubble parameters $\dot{a}=H(z)/(1+z)$, inferred from the different dataset combinations are shown in~\cref{fig:bimodality_comp_adot}, alongside the corresponding CPL predictions. Beyond $z\simeq 0.2$, the expansion histories are largely indistinguishable across models. The most pronounced differences emerge in the redshift interval $z\simeq0.2$--$1.2$, where the six scenarios display the greatest diversity in their expansion behavior. At higher redshifts, CPL generally predicts lower expansion rates. The sensitivity of the models to the expansion dynamics within this intermediate-redshift window suggests that their relative performance is strongly influenced by how accurately they capture the evolution during this phase of slowed expansion.

To summarize, splitting the base chain at $w_a=0$ cleanly isolates the two modes and simultaneously removes the apparent bimodality in the remaining parameters (\cref{fig:bimodal_split_tri} and~\cref{tab:bimodal_split}). While the two branches lead to distinct late-time parameter shifts, most notably in $H_0$ and $\Omega_{\rm m0}$, they share an identical early-time behavior. In both cases, $w(z)\to -1$ at high redshift (\cref{fig:bimodal_split_evol} (top)), reflecting the fact that the CPL$\to -\Lambda$ construction enforces a vacuum-like phase prior to $a_{\rm c}$, with the pre-transition energy density fixed to a constant negative plateau. 
Consequently, the observed bimodality should not be interpreted as evidence for two qualitatively different early-time cosmological histories. Rather, it reflects the likelihood exploring two comparable ways of positioning the NECB-triggered transition while remaining close to a $\Lambda$-like evolution away from the switching epoch.

\section{Conclusions}
\label{sec:conclusions}

In this work, we have examined whether the apparent preference for phantom behavior inferred from current data (traditionally identified through an EoS crossing of $w_{\rm DE}=-1$) persists when DE is described in its most general form, allowing for sign changes in the energy density and hence NECB-crossing. Our goal was deliberately narrow: to test whether the dynamical EoS behavior favored by DESI BAO remains necessary once alternative realizations of phantom behavior are permitted. To this end, we adopted a generalized definition of phantom evolution and introduced two phenomenological models that allow the DE energy density to switch sign, remaining positive today while becoming negative in the past.
The motivation for considering a negative DE density phase is informed by scenarios in which such a transition occurs at intermediate redshifts, comparable to those where the expansion rate $\dot{a}$ is observed to suppressed. In particular, models such as $\Lambda_{\rm s}$CDM exhibit a spontaneous sign switch in the DE density around $z\sim2$, preceding the epoch $z\simeq0.5$--$1$ where the expansion history deviates most strongly from a pure $\Lambda$ behavior.

We constructed CPL$\to-\Lambda$~(\cref{subsec:modelA}) inspired by the work of Ref.~\cite{Ozulker:2025ehg}, where the CPL$_{>a_{\rm c}}$ model was introduced to assess whether NECB-crossing of the EoS is a reconstruction artifact of the linear CPL parametrization or an effective physical feature. We adopted CPL$_{>a_{\rm c}}$ as a control case, retaining its prescription of a cosmological constant phase beyond the NECB-crossing at a redshift implicitly determined by $(w_0,w_a)$, but replacing the positive cosmological constant with a negative one. Complementarily, in sCPL~(\cref{subsec:modelB}), we decoupled the location of the sign switch from NECB-crossing by introducing an explicit transition redshift and allowing the EoS to cross the NECB freely. Together, these constructions enabled us to disentangle which model ingredients are actually favored by the data, within the set of four models considered: CPL, CPL$_{>a_{\rm c}}$, CPL$\to-\Lambda$, and sCPL.

The results convey a clear message: DESI BAO data do not permit negative DE densities within the redshift range they probe. In both sign-switching models, the redshift parameter determining the transition is driven beyond the effective redshift of the highest-$z$ DESI BAO measurement ($z=2.33$, corresponding to $a=0.300$). In sCPL, this manifests as a lower bound on the free parameter $z_\dagger$, while in CPL$\to-\Lambda$ it appears through the derived NECB-crossing redshift $z_{\rm c}$. Consequently, sCPL effectively reduces to CPL from a data analysis perspective, with nearly indistinguishable parameter constraints.
In contrast, enforcing $a_{\rm c} = 1+(1+w_0)/w_a > 0.300$ in the CPL$\to-\Lambda$ model has nontrivial implications for the $(w_0,w_a)$ posteriors. This requirement is effectively equivalent to imposing $(1+w_0)/w_a < -0.700$. A similar constraint arises in Planck\,+\,SNeIa-only analyses, where the transition is pushed to $z_{\rm c} \gtrsim 1.7$, corresponding to $(1+w_0)/w_a \lesssim -0.630$. These additional practical restrictions, together with the requirement $w_a\neq0$ inherent to the CPL$\to-\Lambda$ construction, lead to a significant tightening of the $(w_0,w_a)$ constraints while leaving $(\Omega_{\rm m0}, H_0)$ largely unaffected.

Moreover, the presence of a negative DE density phase in the CPL$\to -\Lambda$ model naturally drives the EoS parameters toward a less dynamical regime. This effect is most pronounced for the base (Planck\,+\,DESI) combination, where the 68\% confidence contours clearly encompass the cosmological constant limits ($w_0\to -1$ and $w_a\to 0$). Nevertheless, owing to the simultaneous tightening of the EoS constraints, the statistical preference for a dynamical DE component does not vanish, with the notable exception of the base case.
From the bimodal structure observed in the base posteriors, we identify an important limitation of the CPL$\to -\Lambda$ construction and, more generally, of model building within the baseline CPL EoS framework. Any model that explicitly ties the onset of NECB-crossing to its location in redshift space (through a parameter such as $z_{\rm c}$) inevitably disfavors behavior arbitrarily close to a cosmological constant, since it requires $w_a\neq 0$ by construction. We further emphasize that DESI BAO data, even when combined with Planck, are insufficient on their own to fully break degeneracies in the EoS parameters, highlighting the essential role of SNeIa data in constraining late-time DE properties.

Overall, none of the models explored in this work outperform CPL, as indicated by their consistently positive $\Delta$AIC values. Nor do they provide a compelling pathway toward resolving the $H_0$ tension. Several factors contribute to these outcomes, including the effective penalization of near-$\Lambda$ behavior in the CPL$\to -\Lambda$ model and the presence of an additional free parameter in sCPL. Nonetheless, within the scope of the scenarios considered here, the dynamical CPL parametrization remains the most effective phenomenological framework for capturing the phantom-like behavior suggested by current cosmological observations.

\section*{Acknowledgments}

The authors are grateful to Emre \"{O}z\"{u}lker and Luis Escamilla for their insightful comments and valuable suggestions. M.G. acknowledges partial support from Scientific and Technological Research Counsil of T\"{u}rkiye (T\"{U}B\.{I}TAK) through a fellowship associated with Grant No.~124N627. The authors thank T\"{U}B\.{I}TAK for their support. \"{O}.A. acknowledges the support by the Turkish Academy of Sciences in the scheme of the Outstanding Young Scientist Award  (T\"{U}BA-GEB\.{I}P). E.D.V. is supported by a Royal Society Dorothy Hodgkin Research Fellowship. This article is based upon work from COST Action CA21136 Addressing observational tensions in cosmology with systematics and fundamental physics (CosmoVerse) supported by COST (European Cooperation in Science and Technology). The authors acknowledge the use of High-Performance Computing resources from the IT Services at the University of Sheffield.\\

\appendix
\section{On the interpretation of one- and two-dimensional parameter constraints}
\label{sec:appendix}

\begin{figure}[h!]
    \centering
    \includegraphics[width=0.96\columnwidth]{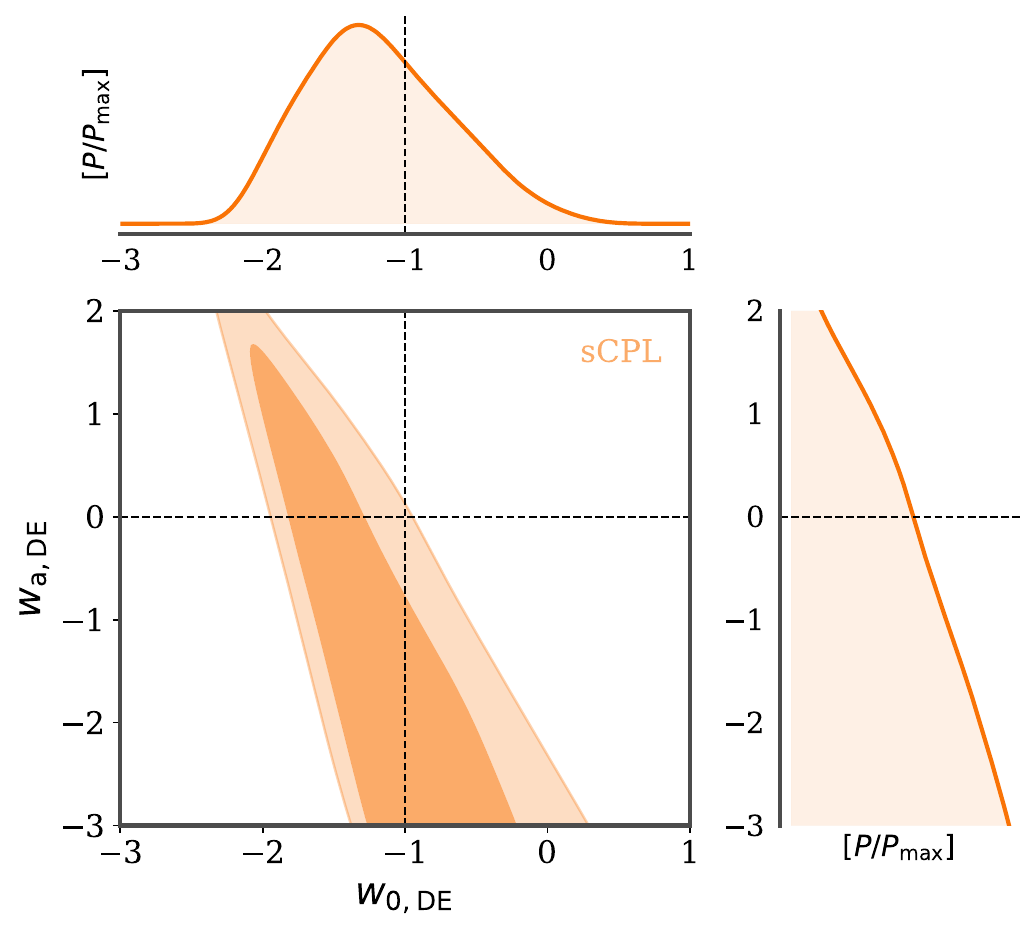}
    \caption{The projection effect on the measurement significances $N_\Lambda$ illustrated using marginalized one- and two-dimensional posteriors of the $w_0$ and $w_a$ parameters from Planck 2018 analyses.\protect\footnotemark[1]}
    \label{fig:projection}
    \footnotetext[1]{\href{https://github.com/williamgiare/wgcosmo/tree/main}{github.com/williamgiare/wgcosmo/tree/main}}
\end{figure}

It is useful to clarify a well-known but sometimes counterintuitive feature of multi-parameter inference, which plays a role in the interpretation of our results. In particular, it is possible for two parameters to be individually consistent with a reference value at the one-dimensional level, while their joint constraints exclude that same reference point in the two-dimensional parameter space. This effect is purely geometric and arises from the projection of a correlated posterior distribution.
In a multi-parameter analysis, the marginalized one-dimensional posterior for a parameter is obtained by integrating the full posterior over all other parameters. As a result, one-dimensional confidence intervals reflect agreement with a reference value after averaging over degeneracies with other parameters. By contrast, two-dimensional confidence regions encode information about correlations between parameters and therefore test whether a specific \emph{pair} of values is jointly compatible with the data.
When parameters are correlated, the posterior probability density can be elongated along a degeneracy direction. In such cases, a reference point may lie well within the one-dimensional marginalized intervals of each parameter separately, yet fall outside the high-probability region of their joint distribution. This situation does not represent an inconsistency or tension in the data, but rather reflects the fact that the data constrain a particular \emph{combination} of parameters more tightly than the individual parameters themselves.
In practical terms, this means that agreement with a reference model should not be assessed solely from one-dimensional constraints. A full assessment requires examining the joint posterior, especially when correlations are present. Throughout this work, we therefore report both one-dimensional significances and two-dimensional joint significances, and we emphasize that apparent agreement at the level of individual parameters does not guarantee compatibility in the full parameter space. \cref{fig:projection} provides a concrete illustration of this projection effect using the sCPL model constrained by Planck-only data. In this case, the one-dimensional constraints show only mild deviations from a cosmological constant, with tensions of $0.4\,\sigma$ in $w_0$ and $0.7\,\sigma$ in $w_a$, while the joint two-dimensional constraint excludes the $(w_0,w_a)=(-1,0)$ point at the $2.3\,\sigma$ level. This discrepancy arises purely from parameter correlations and does not indicate an inconsistency between the one- and two-dimensional results.

\newpage

\bibliography{references}

\end{document}